\documentclass[
  reprint,               
  amsmath,amssymb,       
  aps,                   
  prb,            
  longbibliography       
]{revtex4-2}

\usepackage[english]{babel}
\usepackage{amsmath,amssymb,mathrsfs,mathtools,braket,amsfonts,dsfont}
\usepackage{graphicx}
\usepackage[colorlinks,citecolor=blue,urlcolor=blue,linkcolor=blue]{hyperref}
\usepackage[dvipsnames]{xcolor}
\usepackage{bbm}
\usepackage[T1]{fontenc}
\usepackage{mathtools}

\newcommand{\id}{\mathbbm{1}} 
\renewcommand{\i}{\mathrm{i}}
\newcommand{\e}{\mathrm{e}}

\newcommand{\Tr}{\mathop{\mathrm{Tr}}}

\renewcommand{\Re}{\mathop{\mathrm{Re}}}
\renewcommand{\Im}{\mathop{\mathrm{Im}}}

\begin{document}

\title{Characterizing topology at nonzero temperature:\\
Topological invariants and indicators in the extended SSH model}

\author{Julia D. Hannukainen}
\author{Nigel R. Cooper}
\affiliation{T.C.M. Group, Cavendish Laboratory, J.J. Thomson Avenue, Cambridge CB3 0US, United Kingdom}
\date{\today}

\begin{abstract}
We compare three complementary diagnostics for mixed Gaussian states at nonzero temperature, focusing on the Su-Schrieffer-Heeger (SSH) chain and its inversion-symmetric extension.
Whilst the ensemble geometric phase, a mixed-state generalization of the Zak phase, remains well defined at nonzero temperature, the modulus of the corresponding expectation value vanishes in the thermodynamic limit, limiting its practical use.
To develop diagnostics suitable for large systems, we introduce local twist operators acting on neighboring sites, whose expectation values provide local indicators of the underlying topological phase.
The topological phase is identified from the relative magnitude of these expectation values, which only requires measuring two local expectation values at nonzero temperature, together with one additional nonlocal expectation value when next-nearest-neighbor hopping is included.
In addition, we generalize the local chiral marker to mixed Gaussian states, fully determined by its single-particle correlation matrix, with a nonzero purity gap in their effective single-particle Hamiltonian.
The presence of a purity gap ensures that the correlation matrix can be flattened to an effective projector.
Evaluating the chiral marker with respect to the band-flattened correlation matrix yields a real-space topological invariant that coincides with the winding number in the zero-temperature limit.
The ensemble geometric phase, the local twist operators, and the local chiral marker provide complementary methods to characterize topology in the SSH chain beyond pure states.
\end{abstract}

\maketitle

\section{Introduction}

In equilibrium, topology is a property of the many-body state at zero temperature, which for Gaussian states is the ground state of a quadratic Hamiltonian.
A topological phase can be protected by symmetries, which may be local or spatial in nature.
This means that ground states belonging to distinct topological phases cannot be connected by a finite-depth local unitary circuit that respects the protecting symmetry~\cite{Chen2010}. 
From the Hamiltonian perspective, this implies that parent Hamiltonians of different phases cannot be smoothly deformed into one another without either closing the energy gap or breaking the protecting symmetry~\cite{Chen2010}.
Gaussian states protected by local symmetries are classified~\cite{Schnyder2008,Kitaev2009,Ryu2010,Ludwig2015} into ten symmetry classes~\cite{Cartan1926,Zirnbauer1996,Altland1997}, determined by the combinations of time-reversal, particle-hole, and chiral symmetry.
The distinct phases are characterized by known topological invariants such as the Chern number~\cite{Thouless1982,Kohmoto1985,Read2000}, and winding number~\cite{Ryu2010}, which are evaluated for the ground state wave function.

Quantum systems described by mixed states, arising for example at nonzero temperature or under dissipation, require a different framework for topological classification and characterization.
In this work we adopt a state-based viewpoint, in which topology is defined directly at the level of the density matrix~\cite{Diehl2011,Bardyn2013,Rivas2013,Budich2015,Budich2015_2,Iemini2016,Bardyn2018,Tonielli2020}.
Such mixed Gaussian states can occur as thermal states, and as nonequilibrium steady states of open-system dynamics~\cite{Lindblad1976, Gorini1976, Gardiner1985}.
A general classification of density-matrix topology was introduced in Ref.~\cite{Altland2021}, based on the symmetry properties of the dynamical generators of fermionic matter in and out of equilibrium.
In this formalism, the classification of Gaussian steady states $\rho\sim \e^{-G}$ is formulated in terms of an effective single-particle operator $G$, which plays a role analogous to a quadratic Hamiltonian in equilibrium.
The spectrum of $G$ defines the purity spectrum, which reflects the degree of mixedness of single-particle modes~\cite{Diehl2011,Bardyn2013,Budich2015_2}. 
Eigenvalues far from zero correspond to modes that are predominantly occupied or empty, while eigenvalues at zero represent totally mixed modes.
A finite gap around zero---the purity gap---therefore guarantees that no mode is maximally mixed, and all modes are biased toward being either filled or empty.
The negative-energy eigenstates of the effective Hamiltonian form a Slater determinant that serves as an effective ground state, enabling a topological classification of the corresponding effective Hamiltonian~\citep{Altland2021}.
In the purity-gap framework, topology is defined at the level of the mixed Gaussian state rather than through the finite-temperature spectrum of the physical Hamiltonian~\cite{Diehl2011,Bardyn2013,Altland2021}.
Accordingly, a conventional bulk--boundary correspondence in terms of protected in-gap edge states of $H$ is not evident at nonzero temperature.
Instead, boundary signatures of mixed-state topology has been formulated in terms of the response of the density matrix itself~\cite{Huang2025Interaction}.

Several approaches have been proposed to characterize the topology of mixed states.
The Uhlmann phase~\cite{Uhlmann1976,Viyuela2014,Viyuela2014_2,Huang2014,Budich2015_2} provides a mixed-state generalization of the geometric phase~\cite{Berry1984,Simon1983,Wilczek1984},  by defining parallel transport of density matrices through their purifications in an enlarged Hilbert space. 
However, because every density matrix admits a unique and smooth purification, one can always define a global gauge, making the Uhlmann phase always topologically trivial, even for topologically nontrivial pure states ~\cite{Budich2015_2}.
In contrast, the ensemble geometric phase captures topological properties directly from the density matrix and reduces to the Zak phase, the one-dimensional Berry phase of the occupied band~\cite{Zak1989}, in the pure-state limit~\cite{Bardyn2018}.
The ensemble geometric phase applies to Gaussian states with a nonzero purity gap and has been extended to higher-dimensional systems, interacting states~\cite{Wawer2021,Wawer2021_2,Huang2022}, and a broader framework of topological order parameters~\cite{Huang2025}.
It is constructed from Resta’s formulation of electric polarization, in which the polarization of a one-dimensional insulator is defined as the phase of the expectation value of a large-gauge, many-body twist operator evaluated in the many-body ground state~\cite{Resta1998}.
The operator is defined as $T=\exp[\i(2\pi/L)\sum_j x_j n_j]$, where $L$ is the length of the one dimensional chain, $x_j$ is the position at unit cell $j$, and $n_j$ is the occupation number~\cite{Resta1998}.
For a translation-invariant noninteracting system at zero temperature, Resta’s expression reduces to the Zak phase, connecting the many-body operator to the corresponding single-particle Berry phase.
At nonzero temperature, evaluating the same expectation value with respect to the thermal density matrix yields the ensemble geometric phase~\cite{Bardyn2018}.

The ensemble geometric phase  is well defined for any nonzero temperature and approaches the Zak phase of the lowest band in the purity spectrum in the thermodynamic limit, with finite-size corrections that vanish as the system size increases~\cite{Bardyn2018}.
While the ensemble geometric phase---the phase of the expectation value of the many-body twist operator---remains meaningful at all temperatures, the modulus of the expectation value vanishes exponentially with system size at nonzero temperature, as discussed in Ref.~[\citenum{Molignini2023}].
This is an undesirable feature for a signature of a bulk system and introduces an ambiguity in the practical use of the ensemble geometric phase for large systems.
Such behavior is consistent with general considerations of dipole moment operators, such as the large-gauge translation operator used in Resta’s formulation of polarization, which can exhibit vanishing expectation values in the thermodynamic limit while their phase remains physically meaningful~\cite{Tada2023}.
To examine the vanishing modulus, we focus on the one-dimensional Su–Schrieffer–Heeger (SSH) model~\cite{Su1979,Heeger1988}, which has a quantized Zak phase that takes values of either $0$ or $\pi$ determining the topological phase~\cite{Cooper2019}.
The ensemble geometric phase therefore serves as a nonzero-temperature generalization of the zero-temperature topological invariant of the SSH chain.
We analyze how the modulus of $\langle T\rangle$ depends on temperature and system size, and demonstrate analytically that it decays exponentially with system size at any nonzero temperature.

The exponential decay of the modulus with system size limits the practical evaluation of the ensemble geometric phase in large systems.
To address this, we develop alternative approaches for characterizing topology in inversion-symmetric one-dimensional chains, focusing on the SSH model with both nearest- and next-nearest-neighbor hopping~\cite{Li2014, Zhang2017, Chao2019}.
The SSH model with only nearest-neighbor hopping consists of alternating hopping amplitudes connecting sites within a unit cell (intracell) and between neighboring unit cells (intercell), with the topological phase determined by their relative strength.
We introduce two local twist operators, $T_{j}^{\rm intra}$ and $T_{j}^{\rm inter}$, which reduce $T$ to operators acting on only two sites---within a unit cell and between adjacent unit cells, respectively.
The moduli of the expectation values of the two local operators are periodic, with one period across the length of the chain, and with minima occurring at the center of the chain.
We show that the minimum modulus of $\langle T_{j}^{\rm intra} \rangle$ exceeds that of $\langle T_{j}^{\rm inter} \rangle$ in the trivial phase, while the opposite holds in the topological phase.
Comparing these two minimum values reveals the topological phase from measurements of only two local expectation values.

Extending the SSH model to include next-nearest-neighbor hopping that preserves inversion symmetry introduces an additional topological sector~\cite{Li2014, Chao2019}.
To capture this, we introduce a third local operator,  $T_j^{\rm{nnn}}$, corresponding to the next-nearest-neighbor bond. 
This requires measuring three expectation values---one for each operator---and identifying the topological phase by comparing the relative magnitudes of their minimum moduli.
Since the three operators depend only on local occupation numbers, this method requires only local measurements rather than full knowledge of the global many-body state.
This makes the approach experimentally accessible and scalable to larger systems, offering a practical route to characterize topology at nonzero temperature.

As an additional tool for characterizing the topology of the SSH chain at nonzero temperature, we apply the local chiral marker~\cite{Hannukainen2022} to mixed states with a purity gap. 
In the zero-temperature, translationally invariant limit, the local chiral marker coincides exactly with the chiral winding number, $\nu$, providing a real-space diagnostic of topological phases~\cite{Hannukainen2022,Hannukainen2024}.
For the SSH chain with nearest-neighbor hopping, the winding number $\nu=0,1$ corresponds to the two topologically distinct phases.
Including next-nearest-neighbor hopping introduces a third sector with $\nu=-1$, which cannot be distinguished by the Zak phase, since it remains quantized to $\pi$ for both $\nu=\pm1$.
The chiral marker is defined in terms of the one-particle density matrix, which for Gaussian states at zero temperature, reduces to a projector onto the occupied ground-state bands~\cite{Penrose1956,Koch2001,Bera2015,Bera2017,Kells2018}.
To characterize topology at nonzero temperature, we evaluate the local chiral marker for mixed Gaussian states with a purity gap in its effective Hamiltonian.
In this setting, the density matrix is fully determined by its single-particle correlation matrix, which is related to the effective Hamiltonian through the occupation probabilities of its eigenmodes.
When the purity spectrum is gapped, the correlation matrix can be adiabatically flattened to an effective projector onto the occupied modes, in direct analogy with the band projectors of pure states.
The chiral marker, evaluated with respect to the band flattened correlation matrix, defines the topological phase of the state at nonzero temperature.

For the SSH model, the topology diagnosed by the ensemble geometric phase and the local chiral marker is protected by chiral symmetry.
The ensemble geometric phase approaches the Zak phase of the lowest purity band in the thermodynamic limit; in the translation-invariant SSH model, the quantization of this Zak phase is protected by chiral symmetry.
The local chiral marker is the corresponding real-space topological invariant for states characterized by a $\mathbb{Z}$-valued winding-number classification.
The local twist operators, by contrast, are introduced for inversion-symmetric chains, where inversion symmetry quantizes the polarization and constrains the Wannier-center positions.
This allows shifts of charge centers between topological phases to be inferred from local occupation-based observables.
We compare the ensemble geometric phase, the local twist operators, and the local chiral marker, and show when they are valid, and how they provide complementary ways of characterizing topology in mixed Gaussian states.

\section{Background}
To make this work self-contained and to fix notation, we review the Su–Schrieffer–Heeger model with nearest- and next-nearest-neighbor hopping, and summarise the classification of mixed Gaussian states at nonzero temperature possessing a purity gap.
 
\subsection{The SSH model}
\label{sec:SSH}

The Su–Schrieffer–Heeger (SSH) model is a minimal one-dimensional two-band tight-binding model that captures the essential features of a topological insulator 
protected by inversion symmetry~\cite{Su1979,Heeger1988,Cooper2019}.
Because of its simplicity and well-understood topological structure, it serves as an ideal framework for analyzing how topology can be characterized in mixed Gaussian states at nonzero temperature. 

\begin{figure}[t!]
    \centering
    \includegraphics[width=\columnwidth]{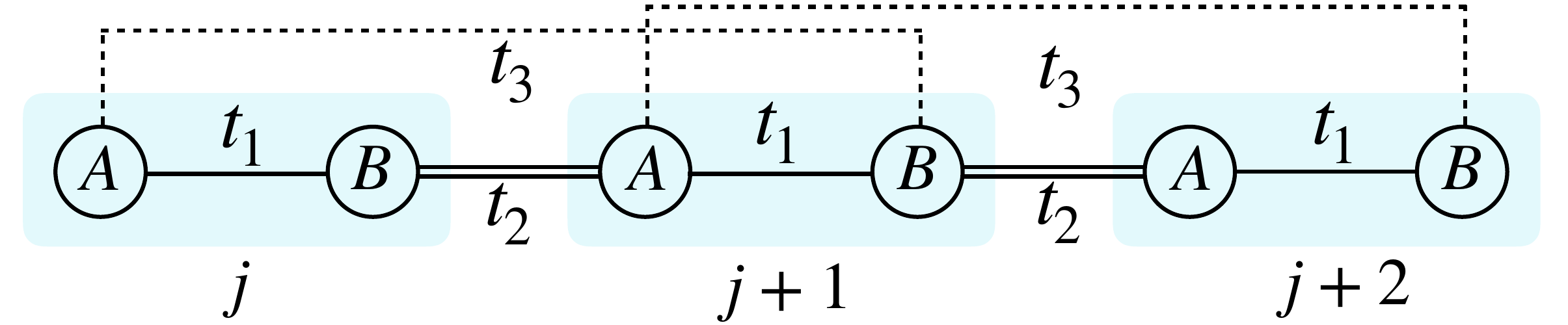}  
    \caption{The SSH model with unit cells labeled by $j$, each containing two sites $A$ and $B$.
  $t_1$ (solid), $t_2$ (double solid), and $t_3$ (dotted) denote the intracell, intercell, and
  next-nearest-neighbor hoppings, respectively.}
    \label{fig:ssh_chain}
\end{figure}

The SSH chain describes spinless fermions with two sublattices, $A$ and $B$, per unit cell and alternating hopping amplitudes connecting sites within and between neighboring unit cells, 
see Fig.~\ref{fig:ssh_chain}.
The model hosts two insulating phases protected by inversion symmetry: a trivial phase, where the intracell hopping dominates, and a topological phase, where the intercell hopping is stronger.
The Hamiltonian with nearest-neighbor hopping reads
\begin{align}
H_{\rm{SSH}} = -\sum_{j=0}^{N-1} \left(t_1\, a^\dagger_j b_j + t_2\, b^\dagger_j a_{j+1} + \mathrm{H.c.}\right),
\label{eq:SSH_Ham}
\end{align}
where $a_j$ ($a^\dagger_j$) and $b_j$ ($b^\dagger_j$) are fermionic annihilation (creation) operators that act on sublattice sites $A_j$ and $B_j$ in the $j$th unit cell.  
The parameters $t_1 \geq 0$ and $t_2 \geq 0$ denote the intracell and intercell hopping strengths, respectively.  
With $N$ unit cells and lattice spacing $a = 1$, the chain has $2N$ sites and length $L = N$.

Assuming a translation invariant lattice with periodic boundary conditions, the SSH Hamiltonian in momentum space becomes
\begin{equation}
H(k) = -\begin{pmatrix}
0 & t_1 + t_2 \e^{-\i k} \\
t_1 + t_2 \e^{\i k} & 0
\end{pmatrix},
\label{eq:HSSH_Momentum}
\end{equation}
where \( k \in [0, 2\pi) \) is the crystal momentum.
Since \(H(k)\) is periodic in \(k\), the Bloch eigenstates define a closed loop as \(k\) traverses the Brillouin zone.
The energy spectrum consists of two bands with eigenvalues $\varepsilon_k = \pm |t_1 + t_2 \e^{\i k}|$, and corresponding normalized eigenstates
\begin{align}
|u_k^\pm\rangle = \frac{1}{\sqrt{2}} \begin{pmatrix}
1 \\
\mp \e^{\i \phi_k}
\end{pmatrix},
\label{eq:egensvecSSH}
\end{align}
where \( \phi_k = \arg(t_1 + t_2 \e^{\i k}) \) is the phase of the off-diagonal matrix element in the Hamiltonian Eq.~\eqref{eq:HSSH_Momentum}.
The Zak phase of the lower band is defined as
\begin{equation}
\varphi_{\rm Zak} = \int_0^{2\pi} \langle u_k^- | \i \partial_k | u_k^- \rangle \, \mathrm{d}k = -\frac{1}{2} \int_0^{2\pi} \frac{\partial \phi_k}{\partial k} \, \mathrm{d}k;
\end{equation}
it is a geometric phase accumulated by a Bloch band as $k$ traverses the Brillouin zone.
In the convention implicit in Eq.~\eqref{eq:HSSH_Momentum}, the Zak phase of the lower band is quantized by inversion symmetry to $0$ for $t_1>t_2$ and $\pi$ for $t_2>t_1$.
A change of Bloch-basis convention associated with a different choice of unit cell can shift the absolute Zak phase by a quantized amount, while the relative phase difference between the trivial and topological regimes remains invariant~\cite{Atala2013,deJuan2014,Ling15}.
Geometrically, as \( k \) runs from \( 0 \) to \( 2\pi \), the complex function \( t_1 + t_2 \e^{\i k} \) traces a closed loop in the complex plane, with \( \phi_k \) the polar angle along this path.
The total change of \(\phi_k\) over the Brillouin zone defines the winding number \(\nu = -\varphi_{\mathrm{Zak}}/\pi\), which changes only at the gap closing \(t_1 = t_2\), distinguishing the trivial (\(\nu = 0\)), and topological (\(\nu = 1\)) phases.

The SSH chain is generalized by including inversion symmetry preserving next-nearest-neighbor-couplings~\cite{Li2014,Chao2019}.
These additional hopping terms connect $A$ sites in one unit cell to $B$ sites in the neighboring cell two sites away,  extending the Hamiltonian in Eq.~\eqref{eq:SSH_Ham} to
\begin{align}
H_{\rm{eSSH}} = &-\sum_{j=0}^{N-1} \big(
t_1\, a^\dagger_j b_j 
+ t_2\, b^\dagger_j a_{j+1} 
+ t_3\, a^\dagger_j b_{j+1} 
+ \mathrm{H.c.} \big),
\label{eq:H_SSH_nnn}
\end{align}
where $t_3\geq 0$ denotes the next-nearest-neighbor hopping strength.
For a translation invariant lattice the Zak phase still takes values of either \(0\) when $t_1>|t_2+t_3|$ or \(\pi\) when $t_1<|t_2+t_3|$,  due to inversion symmetry.
However, the winding number now distinguishes three different topological phases: the trivial phase $\nu=0$ for $t_1>|t_2+t_3|$, and the two topological phases, $\nu=1$ for $t_1<|t_2+t_3|$ where $t_2>t_3$, and $\nu=-1$ for $t_1<|t_2+t_3|$ where $t_2<t_3$.
The topologically nontrivial phases \(\nu = \pm 1\) host zero-energy boundary modes under open boundary conditions; the sign of the winding distinguishes the two bulk topological sectors.

At nonzero temperature the SSH chain is in a thermal mixed state.
Because the SSH Hamiltonian is quadratic, the thermal state remains Gaussian and is fully characterized by single-particle correlations. 
Thermal excitations lead to partial occupation of both bands, and the system is no longer in a pure state. 
As a result, the Zak phase, defined strictly for the ground state is no longer a well defined topological invariant.

\subsection{Topological classification of mixed Gaussian states with a purity gap}

A number-conserving fermionic Gaussian mixed state has the form
\begin{align}
\rho = Z^{-1} e^{-\hat G},
\qquad
\hat G = \sum_{ij} c_i^\dagger G_{ij} c_j,
\end{align}
where $c_i$ and $c_i^\dagger$ are fermionic annihilation and creation operators, and $G$ is the corresponding single-particle Hermitian matrix.
The partition function $Z=\Tr(\e^{-\hat{G}})$ enforces the normalization $\Tr(\rho)=1$, where \(\Tr\) denotes the trace over the many-body Fock space.
In this work we focus on thermal states, for which \(\hat G=\beta \hat H\), and
\begin{align}
\rho = Z^{-1} e^{-\beta \hat H},
\qquad
\hat H = \sum_{ij} c_i^\dagger H_{ij} c_j.
\end{align}
At the single-particle level this implies $G=\beta H$, where \(H\) is the single-particle matrix representation of \(\hat H\), and $\beta$ is the inverse of temperature times Boltzmann's constant.
The Hermitian matrix $G$, referred to as the effective Hamiltonian, determines both the occupations and the purity of the state through its eigenvalues~\cite{Bardyn2013}.
If the spectrum of $G$ is gapped around zero, all modes are either predominantly occupied or predominantly empty, and the state is adiabatically connected to a pure state.
In this case, the topology is fully determined by that corresponding pure state.

The topological classification depends on the structure of this occupied subspace: the eigenstates of $G$ below the purity gap form an occupied subspace, analogous to the occupied band that defines the ground state of a free-fermion Hamiltonian~\cite{Bardyn2013, Altland2021}.
Each eigenvalue $\epsilon_\gamma$ of $G$ defines a fermionic mode with thermal occupation $p_\gamma = [\e^{\epsilon_\gamma}+1]^{-1}$.
The mixedness of mode $\gamma$ is quantified by the purity $\eta_\gamma \equiv 1 - 2 p_\gamma$, for which $|\eta_\gamma|=1$ for a  pure mode and $|\eta_\gamma|=0$ for a maximally mixed mode.
The purity gap refers to a nonzero gap around $\eta_\gamma=0$.
Using $p_\gamma = [\e^{\epsilon_\gamma}+1]^{-1}$, leads to the relation
\begin{equation}
\eta_\gamma 
= \frac{\e^{\epsilon_\gamma}-1}{\e^{\epsilon_\gamma}+1}
= \tanh(\epsilon_\gamma/2),
\end{equation}
showing how the purity spectrum is directly determined by the eigenvalues of $G$.

The occupation probabilities $p_\gamma$ are contained in the two-point correlators,
\begin{align}
&  \langle c_i^\dagger c_j \rangle
  = [f(G)]_{ij},
\end{align}  
where
\begin{align}
  &f(G) = (\id + \e^{G})^{-1}.
  \label{eq:C-from-G}
\end{align}
In the pure-state limit, where $p_\gamma \in \{0,1\}$, the matrix $f(G)$ is a projector onto the ground state, which determines the topological classification.
For a mixed state, $f(G)$ has eigenvalues in the interval $(0,1)$.  
When the purity spectrum is gapped around $p_\gamma = 1/2$, spectral flattening of $f(G)$ yields the band flattened projector
\begin{align}
  P = \frac{1}{2}\left(\frac{2f(G)-1}{|2f(G)-1|}+1\right).
  \label{eq:P-from-fG}
\end{align}
The matrix $P$ is the object entering the topological classification, in direct analogy with the projectors onto the filled bands in the classification of pure Gaussian ground states.
The topology of a mixed Gaussian state with a purity gap is therefore determined by its correlation matrix, or equivalently through the effective Hamiltonian.

\section{The Ensemble Geometric Phase}
The ensemble geometric phase is a topological invariant for one dimensional mixed Gaussian states, $\rho~\sim \e^{-G}$ with a purity gap~\cite{Bardyn2018}.
This phase reduces to the Zak phase of the lowest purity band of the corresponding effective Hamiltonian, $G$ in the thermodynamic limit~\cite{Bardyn2018}.
The ensemble geometric phase is defined as~\cite{Bardyn2018}.
\begin{equation}
\varphi_{\rm{EGP}}= \Im\log\langle T\rangle\,.
\end{equation}
For the SSH model, the many-body translation operator $T$ takes the explicit form
\begin{equation}
T=\exp\!\left\lbrace\i\,\delta k\sum_{j=0}^{N-1}\left[x_j n_j+\delta x(n_j^{B}-n_j^{A})\right]\right\rbrace,
\label{eq:Tperator_full}
\end{equation}
where $\delta k = 2\pi/L$  is the minimal spacing between discrete crystal momenta in the Brillouin zone for a chain of length $L$, \(n^{\rm A}_j = a_j^\dagger a_j \) and \(n^{\rm B}_j =  b_j^\dagger b_j \) are the number operators on sublattice sites \( A \) and \( B \) in the \( j \)th unit cell, and $n_j=n_j^{A}+n_j^{B}$.
$x_j$ is the unit-cell coordinate and the two orbital positions within the cell are $x_j^{A}=x_j-\delta x$ and $x_j^{B}=x_j+\delta x$, such that the intra-cell separation is $2\delta x$.

Assuming translation invariance, the expectation value of $T$ in momentum space for the SSH model at nonzero temperature, takes the form~\cite{Bardyn2018}
\begin{equation}
\langle T \rangle = \frac{1}{Z}\,\det\!\left[\sigma_0 + M_{\mathrm{T}}\right].
\label{eq:T_expvalue}
\end{equation}

Here, $Z = \mathrm{Tr}\!\left(\e^{-\beta H_{\mathrm{SSH}}}\right)$ is the partition function that normalizes the thermal density matrix, 
$\sigma_0$ denotes the two-dimensional identity matrix, and $M_{\mathrm{T}}$ is defined as
\begin{equation}
M_{\mathrm{T}} = (-1)^{N-1} \prod_{k=0}^{N-1} \e^{-\beta E_{k+1}}\, U_{k+1}^\dagger U_k,
\label{eq:M-matrix}
\end{equation}
where $U_k$ is the unitary matrix of eigenvectors of $H_{\mathrm{SSH}}(k)$, and $E_k$ is the diagonal matrix of the corresponding single-particle energies.
The partition function is real,  so the ensemble geometric phase is given by
\begin{align}
\varphi_{\rm{EGP}}=\Im\log\det[\sigma_0+M_{\rm{T}}].
\end{align}
For finite systems, $\varphi_{\rm{EGP}} = \varphi_{\rm{Zak}} + \Delta(N)$, where $\Delta(N)$ is a model-dependent correction that vanishes as $N \to \infty$.
This means that the ensemble geometric phase reduces to the Zak phase $\varphi_{\rm{Zak}}$ of the lowest band of the energy spectrum of $\beta H_{\rm{SSH}}(k)$ in the thermodynamic limit.
The ensemble geometric phase therefore distinguishes the trivial and the topological phases of the SSH model with only nearest-neighbor coupling.
When next-nearest-neighbor coupling is introduced, the ensemble geometric phase---being sensitive only to a phase modulo $2\pi$---cannot tell apart the two topological phases whose zero temperature winding numbers differ by a sign,  $\nu=\pm 1$.

\subsection{The modulus of $\langle T \rangle$ tends to zero in the thermodynamic limit}

\begin{figure}[t!]
    \centering
    \includegraphics[width=\columnwidth]{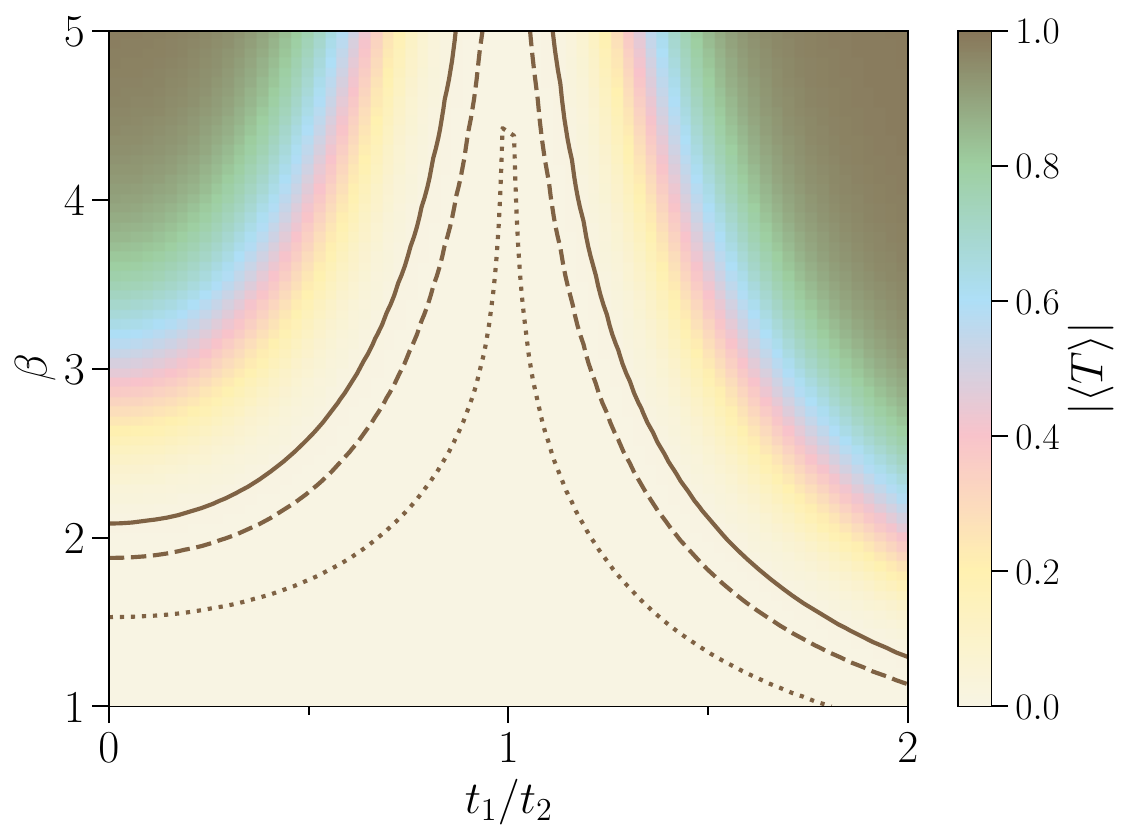}  
    \caption{The amplitude $|\langle T \rangle|$ as a function of $t_1/t_2$ and inverse temperature $\beta$, where $t_1$ is the intracell hopping energy, $t_2=2$ is the intercell hopping energy in the SSH model. The number of unit cells is $N=150$. The overlaid contour lines in brown indicate the loci where $|\langle T \rangle|=0.01$ for systems with $N=150$ (solid), $N=100$ (dashed), and $N=50$ (dotted). The embedding is inversion-symmetric, with unit-cell centers $x_j=j+1/2$, with $j=0,...,N-1$ ($N$ even), and orbital positions $x_j^A=x_j-\delta x$, $x_j^B=x_j+\delta x$ with $\delta x=1/4$.}
    \label{fig:Modulus_Tfull}
\end{figure}

The modulus of the expectation value, $|\langle T \rangle|$, equals one for pure states, but decreases exponentially with increasing system size and temperature for mixed Gaussian states~\cite{Molignini2023}.
Fig.~\ref{fig:Modulus_Tfull} shows the dependence of $|\langle T \rangle|$ on the inverse temperature $\beta$ and the hopping ratio $t_1/t_2$ for the SSH model.
The color map corresponds to a system with $N=150$ unit cells, while the overlaid contour lines indicate the loci where $|\langle T \rangle|=0.01$ for systems with $N=150$ (solid), $N=100$ (dashed), and $N=50$ (dotted).
As the system size increases, the region where $|\langle T \rangle|$ remains nonzero shrinks, signaling that the expectation value vanishes in the thermodynamic limit even though the ensemble geometric phase remains well defined.
Consequently, the ensemble geometric phase provides a meaningful characterization of topology in finite systems, but becomes difficult to measure practically in large systems.

The topologically trivial phase with intercell hopping \( t_2 = t_3 = 0 \) provides a simple example illustrating the vanishing modulus for large system sizes.
The Hamiltonian in Eq.~\eqref{eq:HSSH_Momentum} reduces to \( H(k) = -t_1 \sigma_x \) in the trivial limit, where $\sigma_x$ the Pauli matrix acting in sublattice space.
The corresponding energy spectrum is flat \( \varepsilon_k = \pm t_1 \), with $k$ independent eigenvectors \( |u_k^\pm\rangle = (1, \mp 1)^T / \sqrt{2} \).
In this case, $ U_{k+1}^\dagger U_k = \sigma_0$  for all \( k \), and 
\begin{equation}
|\det(\sigma_0+M_{\rm{T}})|=|1+s\e^{\beta t_1 N}||1+s\e^{-\beta t_1 N}|,
\label{eq:numt1_lim}
\end{equation}
where $s=(-1)^{N-1}$.
The partition function in the trivial limit is
\begin{align}
Z=\left[\left(1 + \e^{\beta t_1} \right) \left(1 + \e^{-\beta t_1}\right)\right]^N,
\label{eq:dent1_lim}
\end{align}
and by simplifying the quotient between Eq.~\eqref{eq:numt1_lim} and Eq.~\eqref{eq:dent1_lim}, the modulus of $\langle
T\rangle$ is
\begin{align}
|\langle T\rangle| = \e^{-2N\ln(1+\e^{-\beta t_1})}.
\end{align}
At zero temperature, in the $\beta\rightarrow\infty$ limit, the modulus goes to one, but at any nonzero temperature the modulus approaches zero in the thermodynamic limit.
The same trend of a vanishing modulus $|\langle T\rangle |$ persists for general finite values of the hopping parameters $t_1$, $t_2$, and $t_3$. 
In the large $N$ limit 
\begin{align}
(U_{k+1}^\dagger U_k)_{\alpha\beta}& =  \langle u_{k + 1, \alpha} | u_{k, \beta} \rangle \simeq 1 - \delta k \langle u_{k, \alpha} \partial_k u_{k, \beta}\rangle\nonumber\\
& \simeq \exp[\i \delta k (A_k)_{\alpha\beta}],
\label{eq:unitaryexp}
\end{align}
where $A_k$ denotes the non-Abelian Berry connection, defined as
\begin{equation}
    (A_k)_{\alpha\beta} = \i \bra{u_{k, \alpha}} \partial_k u_{k, \beta} \rangle.
\end{equation}
To estimate an upper bound on $|\langle T\rangle|$, the matrix $M_{\rm{T}}$, Eq.~\eqref{eq:M-matrix}, is decomposed into diagonal and off-diagonal components, where $\e^{-\beta E_{k}}=\e^{-\beta\varepsilon_k \sigma_3}$, and $\varepsilon_k$ is the absolute eigenenergy of the SSH Hamiltonian at a given momenta $k$.
Following Ref.~[\citenum{Bardyn2018}], Eq.~\eqref{eq:unitaryexp} is parameterized as:
\begin{align}
    U_{k+1}U_k & = \e^{\i \delta k \sum_{i = 0, 3} \mathcal{A}_k^i \sigma_i} + \delta k \sum_{i = 1, 2} \mathcal{A}_k^i \sigma_i + \mathcal{O}(\delta k^2) \nonumber \\
    & \equiv U^\text{diag}_{k+1, k} + V_k,
    \label{eq:U_diag.V}
\end{align}
where $\sigma_i$ for $i\in(1,2,3)$ are the Pauli matrices, $\sigma_0$ the identity matrix in two dimensions, where the components $\mathcal{A}_k^i$ are real.
The diagonal matrix $U^\text{diag}_{k+1, k}= \e^{\i \delta k (\mathcal{A}_k^0 \sigma_0 + \mathcal{A}_k^3 \sigma_3)}$, where $\mathcal{A}_k^0 \pm \mathcal{A}_k^3 = \i \langle u_k^{\pm}| \partial_k u_{k}^{\pm} \rangle$, and the off diagonal matrix $V_k =\delta k (\mathcal{A}_k^1 \sigma_1 + \mathcal{A}_k^2 \sigma_2)$ mixes the two bands.
Expanding the determinant in Eq.~\eqref{eq:T_expvalue} perturbatively in the off-diagonal transition matrices $V_k$  yields:
   \begin{align}
    \det(\sigma_0+M_{\rm{T}})= \det \left(G^{-1} + V^{(1)} + V^{(2)} + \ldots \right),
     \label{eq:perturb_M}
  \end{align}    
  where 
\begin{align}
&G^{-1} = \sigma_0 + W_{N-1, 0},
\end{align}
and $W_{k_1,k_2}$ is diagonal with components
\begin{equation}
W^{\pm}_{k_1,k_2}=\prod_{k_2\leq k\leq k_1}\e^{\i\delta k(\mathcal{A}_k^0\pm \mathcal{A}_k^3)}\e^{\mp\beta\varepsilon_k}.
\end{equation}
The leading terms in the expansion Eq.~\eqref{eq:perturb_M} describe successive interband transition processes:
\begin{align}
&V^{(1)} = \sum_k W_{N-1, k} V_k W_{k-1, 0},\\
& V^{(2)} = \sum_{k > k'} W_{N-1, k} V_k W_{k-1, k'} V_{k'} W_{k'-1, 0},
\end{align}  
where $V^{(1)}$ is off diagonal, and $V^{(2)}$ is diagonal in band space.
For large $N$ the zeroth order contribution in Eq.~\eqref{eq:perturb_M} is dominated by the lower band
\begin{align}
\det G^{-1}\approx \e^{N\beta \bar{\varepsilon} }  \e^{\i \delta k \sum_k(\i \langle u_k^{-}| \partial_k u_{k}^{-} \rangle)},
\end{align}
where we have introduced the average single-particle energy $\bar{\varepsilon} = \frac{1}{N}\sum_k \varepsilon_k$, such that $\sum_k \beta \varepsilon_k \approx N \beta \bar{\varepsilon}$.
So, the modulus of $\det G^{-1}$ grows exponentially with system size.

To evaluate the perturbative contribution to the modulus, it is convenient to take the logarithm of the determinant, which converts matrix products into traces and permits a systematic expansion in \(V_k\):
\begin{align}
\ln \det [\sigma_0+ M_{\rm{T}}] &=  \ln \det G^{-1}+\Delta(N) + \mathcal{O}(\delta k^3),\\
&\Delta(N) = \Tr G V^{(2)} - \tfrac{1}{2} \Tr (G V^{(1)})^2,
\label{eq:V_expansion}
\end{align}
to second order in $V_k$.
Terms linear in \(V_k\) are off-diagonal in band space and therefore do not contribute to the determinant at this order.
A detailed analysis (see Appendix~\ref{app:perturbative_expansion}) shows that the real part of the lowest-order correction decays as
\(\Re[\Delta(N)] \sim N^{-1}\) for large system size, and the modulus of the numerator in Eq.~\eqref{eq:T_expvalue} behaves as
\begin{align}
|\det(\sigma_0+M_{\mathrm{T}})|
\simeq \e^{N\beta\bar{\varepsilon}}\!\left(1 + \mathcal{O}(N^{-1})\right),
\end{align}
for large \(N\).

The partition function in the denominator $\langle T\rangle$ in Eq.~\eqref{eq:T_expvalue} is given by the product over momenta,
\begin{equation}
Z = \prod_{k=0}^{N-1} \left(1 + \e^{-\beta \varepsilon_k}\right)\left(1 + \e^{\beta \varepsilon_k}\right).
\end{equation}
At zero temperature, $|Z| = |\det(\sigma_0 + M_{\rm T})|$ and $|\langle  T  \rangle| = 1$.
For nonzero temperature, the partition function expands as a sum over all subsets of $\{0, \ldots, N{-}1\}$. 
Separating out the contributions from the empty set and the full set yields
\begin{equation}
\prod_{k=0}^{N-1} (1 + \e^{\pm\beta \varepsilon_k}) 
= 1 + \e^{\pm\beta N\overline{\varepsilon}} 
+ \sum_{\mathrlap{\substack{S \subsetneq \{0,\ldots,N{-}1\} \\ S \neq \emptyset}}}
\phantom{\sum} \e^{\pm\beta \sum_{k \in S} \varepsilon_k},
\label{eq:Zinsets}
\end{equation}
where the sum runs over all non-empty, proper subsets $S$.
Equation~\eqref{eq:Zinsets} shows that $Z$ includes exponential terms $\sim \e^{\beta r \overline{\varepsilon}}$ for $1 \leq r \leq N$, many with $r \sim N$, so in the thermodynamic limit it grows faster than the single exponential term in the numerator.
This implies that the modulus $|\langle T\rangle |$ vanishes for large enough $N$.

\section{Local twist operators}

We introduce an alternative indicator of topology in inversion-symmetric chains at nonzero temperature by defining two local twist operators acting on sites within a single unit cell and on adjacent sites across a unit cell boundary.
Unlike the ensemble geometric phase, which is defined from a global many-body twist operator, the local twist operators are motivated by the quantized shift in the spatial localization of electronic charge that distinguishes the topological phases of the SSH model.
Due to inversion symmetry, the electronic charge localizes either on atomic sites or on the bonds between them.
Accordingly, the electric polarisation, $P^{(\pm)}_{\rm{el}}$,  is quantized in terms of the Zak phase: $P^{(\pm)}_{\rm{el}}=e\varphi_{\rm{Zak}}^{\pm}/(2\pi)$, where $e$ is the electric charge~\cite{King-Smith1993,Vanderbilt1993}.
In real space, the polarization is described in terms of Wannier functions constructed from Bloch states via Fourier transformation.
The center of a Wannier function associated with a filled band $n$---the Wannier center $x_n =  \langle w_n | \hat{x} | w_n \rangle$---specifies the real-space position of the corresponding electronic state~\cite{Wannier1937}.
This position is proportional to the Zak phase (up to a lattice constant), and the electric polarization corresponds to the average position of these centers: $P_{\text{el}} = -e \sum_{\text{occ}} x_n$, where the sum runs over occupied bands~\cite{King-Smith1993,Vanderbilt1993}.
In the trivial phase, dominated by intracell hopping, the Wannier centers lie on the atomic sites and the polarization vanishes. 
In the topological phase, dominated by intercell hopping, the centers shift by half a unit cell onto the bonds, yielding $P_{\rm{el}}=e/2$.

\begin{figure}[t!]
    \centering
    \includegraphics[width=1\linewidth]{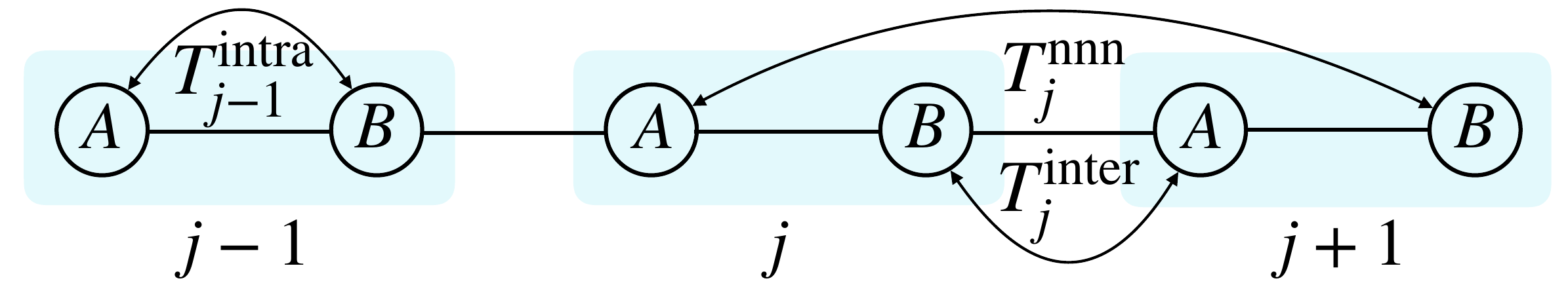}  
    \caption{The SSH chains with three cells labeled by $j$, depicting the local intracell operator $T_{j-1}^{\rm{intra}}$, the local inter cell operator $T_{j}^{\rm{inter}}$,  and the nonlocal next-nearest-neighbor operator $T_{j}^{\rm{nnn}}$.}
    \label{fig:operators}
\end{figure}

To capture the difference in the charge localisation between the two topological phases, we define the two operators
\begin{align}
T_j^{\rm intra}
&=\exp\left\lbrace\i\,\delta k\,[x_j n_j+\delta x(n_j^{B}-n_j^{A})]\right\rbrace,
\label{eq:Tintra}
\end{align}
which acts on the two sites of unit cell $j$, and
\begin{align}
T_j^{\rm inter}
&=\exp\left\lbrace\i\,\delta k\,[x_j n_j^{B}+x_{j+1} n_{j+1}^{A}+\delta x(n_j^{B}-n_{j+1}^{A})]\right\rbrace,
\label{eq:Tinter}
\end{align}
which acts on the bond connecting sites $B_j$ and $A_{j+1}$ in adjacent cells, see Fig.~\ref{fig:operators}.
These operators are constructed such that the product over all local contributions reproduces the full many-body translation operator $T$, Eq.~\eqref{eq:Tperator_full}, whose expectation value defines the ensemble geometric phase.
The local form of Eq.~\eqref{eq:Tintra}, and Eq.~\eqref{eq:Tinter} reflects a decomposition of $T$ into factors acting on pairs of neighboring sites, corresponding to charge displacements either within a unit cell or across a unit cell boundary. 
This construction motivates the question of whether the global ensemble geometric phase can be reconstructed from the phases of local expectation values; in other words, whether
\begin{align}
\varphi^{\ell}_{\mathrm{EGP}}
&=\Im\ln\,\langle T\rangle
=\Im\ln\!\big\langle\prod_{j=0}^{N-1}T_j^{\ell}\big\rangle,
\label{eq:globalEGP}
\end{align}
for $\ell\in\{\mathrm{intra},\mathrm{inter}\}$, coincides with the total phase obtained from individual expectation values,
\begin{align}
\varphi^{\ell}_{\mathrm{tot}}
&=\Im\ln\!\prod_{j=0}^{N-1}\langle T_j^{\ell}\rangle
=\sum_{j=0}^{N-1}\Im\ln\,\langle T_j^{\ell}\rangle
=\sum_{j=0}^{N-1}\varphi^{\ell}_j.
\label{eq:localphase_sum}
\end{align}
In general, the expectation value of a product of local operators does not factorize into the product of their expectation values, since hopping terms in the Hamiltonian couple neighboring sites and generate spatial correlations.
Comparing Eqs.~\eqref{eq:globalEGP} and~\eqref{eq:localphase_sum} allows us to assess whether, and under what conditions, the global geometric phase can be reconstructed from spatially local observables, and thereby to identify local signatures of the topological phase.

$T_j^{\rm intra}$ and $T_j^{\rm inter}$ are diagonal in the single-particle site basis and equal to the identity away from the two orbitals they act on. 
The nontrivial diagonal entries of $T_j^{\rm intra}$ are $\exp[\i\,\delta k (x_j-\delta x)]$ on $A_j$ and $\exp[\i\,\delta k (x_j+\delta x)]$ on $B_j$. 
For $T_j^{\rm inter}$ the nontrivial entries are $\exp[\i\,\delta k (x_j+\delta x)]$ on $B_j$ and $\exp[\i\,\delta k (x_{j+1}-\delta x)]$ on $A_{j+1}$.

The expectation values of the twist operators, Eq.~\eqref{eq:Tintra}, and Eq.~\eqref{eq:Tinter}, are computed in real space using the definition
\begin{align}
\langle O\rangle=\frac{1}{Z}\det[\id_{2N}+\e^{-\beta H_{\rm{SSH}}}O],
\label{eq:determinant_exp_expression}
\end{align}
for $O\in\{T_j^{\rm intra},\,T_j^{\rm inter}\}$, where $H_{\rm SSH}$ is the $2N\times 2N$ single-particle, first-quantized, SSH Hamiltonian.
The corresponding partition function is defined as
\begin{align}
Z=\det[\id_{2N}+\e^{-\beta H_{\rm{SSH}}}].
\end{align}
The determinant expression in Eq.~\eqref{eq:determinant_exp_expression} is useful for analytical purposes but becomes numerically unstable for large $N$ and $\beta$.
A numerically stable form is obtained by dividing Eq.~\eqref{eq:determinant_exp_expression} by the partition function $Z$ and rewriting the result in terms of the single-particle 
correlation matrix, 
\begin{align}
\langle c_i^\dagger c_j\rangle 
    = [f(H_{\rm SSH})]_{ij} 
    = \big[\id_{2N} + \e^{\beta H_{\rm SSH}}\big]^{-1}_{ij}.
\end{align}
By using the identity $(1+\e^{-\beta H_{\rm SSH}})^{-1} = 1 - f(H_{\rm SSH})$, the expectation value in Eq.~\eqref{eq:determinant_exp_expression} is rewritten as
\begin{align}
\langle O\rangle &= \det\!\big[\id_{2N} - f(H_{\rm SSH}) + f(H_{\rm SSH})\,O\big].
\label{eq:Levitov}
\end{align}

The expectation values $\langle T_j^{\mathrm{intra}} \rangle$ and $\langle T_j^{\mathrm{inter}} \rangle$ take simple closed analytical forms in the strict trivial  ($t_1>0$, $t_2=0$) and strict topological ($t_1=0$, $t_2>0$) limits.
The expectation values of the intracell operator in the trivial and topological limits, for periodic boundary conditions, are
\begin{align}
\langle T_j^{\mathrm{intra}} \rangle_{\rm triv} 
&= \frac{ 1 + 2 \cosh(\beta t_1) \cos(\delta k \, \delta x)\, \e^{\i \delta k x_j} + \e^{2 \i \delta k x_j} }
{2 [1 + \cosh(\beta t_1)]}, \label{eq:Tintra-triv} \\
\langle T_j^{\mathrm{intra}} \rangle_{\rm top} 
&=\frac{(1+\e^{\i\delta k(x_j-\delta x)})(1+\e^{\i\delta k(x_j+\delta x)})}{4}.\label{eq:Tintra-top}
\end{align}
The corresponding expectation values of intracell operators take the form
\begin{align}
\langle T_j^{\mathrm{inter}}\rangle_{\rm triv} 
&=\frac{(1+\e^{\i\delta k(x_{j+1}-\delta x)})(1+\e^{\i\delta k(x_j+\delta x)})}{4}, \label{eq:Tinter-triv}\\
\langle T_j^{\mathrm{inter}} \rangle_{\rm top} 
&= \frac{1 + 2 \cosh(\beta t_2)\cos(\delta k \delta x)\, \e^{\i \delta k x_j^{\rm bond}}}{2[1+\cosh(\beta t_2)]} \nonumber\\
&\quad + \frac{\e^{2 \i \delta k x_j^{\rm bond}}}{2[1+\cosh(\beta t_2)]},\label{eq:Tinter-top}
\end{align}
where $x_j^{\rm bond}=(x_j+x_{j+1})/2$ is the bond-center coordinate.
The SSH Hamiltonian is the same for both periodic and open boundary conditions in the strict trivial limit,  so the expectation values in Eq.~\eqref{eq:Tintra-triv}, and Eq.~\eqref{eq:Tinter-triv}, are the same for both choices of boundary conditions.
In the topological limit the choice of boundary conditions only affects the expectation value of $T_{N-1}^{\mathrm{inter}} $ which connects the two boundaries nonlocally, where 
\begin{align}
\langle T_{N-1}^{\mathrm{inter}} \rangle_{\rm top}^{\rm{OBC}}
&=\frac{(1+\e^{\i\delta k(x_0-\delta x)})(1+\e^{\i\delta k(x_{N-1}+\delta x)})}{4}.
\end{align}
The expectation values $\langle T_j^{\mathrm{intra}} \rangle_{\rm top}$, and $\langle T_j^{\mathrm{inter}}\rangle_{\rm triv} $ are independent of temperature: The $\beta$ dependence in the numerator of these expectation values is factorised, and cancelled by the partition function in the denominator.
This suggests that only $\langle T_j^{\mathrm{intra}} \rangle_{\rm triv}$, and $\langle T_j^{\mathrm{inter}} \rangle_{\rm top}$ are relevant expectation values for characterizing topology at a nonzero temperature.
The expectation value of the full translation operator is equal to the product of the local twist operators in the strict limits, where
\begin{align}
\langle T\rangle_{\rm{triv}} &=\prod_{j=0}^{N-1}\langle T_j^{\rm{inter}}\rangle_{\rm{triv}},\\
\langle T\rangle_{\rm{top}}& =\prod_{j=0}^{N-1}\langle T_j^{\rm{intra}}\rangle_{\rm{top}}.
\end{align}
This means that the ensemble geometric phase in these two limits is equal to the sum of the phases of the local operators across the chain.
For $t_2=0$:
\begin{equation}
\begin{aligned}
\varphi_{\rm{EGP}}&=\Im \ln\langle T\rangle=\sum_{j=0}^{N-1}\Im \ln \langle T_j^{\rm{inter}}\rangle_{\rm{triv}}
=\sum_{j=0}^{N-1} \varphi_j^{\rm{inter}},
\label{eq_egp_sum_triv}
\end{aligned}
\end{equation}
and for $t_1=0$:
\begin{equation}
\begin{aligned}
\varphi_{\rm{EGP}}&=\Im \ln \langle T\rangle=\sum_{j=0}^{N-1} \Im \ln\langle T_j^{\rm{intra}}\rangle_{\rm{top}}
=\sum_{j=0}^{N-1} \varphi_j^{\rm{intra}}.
\label{eq_egp_sum_top}
\end{aligned}
\end{equation}
The ensemble geometric phase reduces to a sum of local phases in the strict limits,  and the same structure extends to the full parameter space, but only when the intracell operator is used in the limit  $t_1>t_2$, and the intercell operator is used in the limit $t_1<t_2$.
We demonstrate this by using the inversion symmetry of the SSH chain.

\begin{figure*}
    \centering
    \includegraphics[width=1\textwidth]{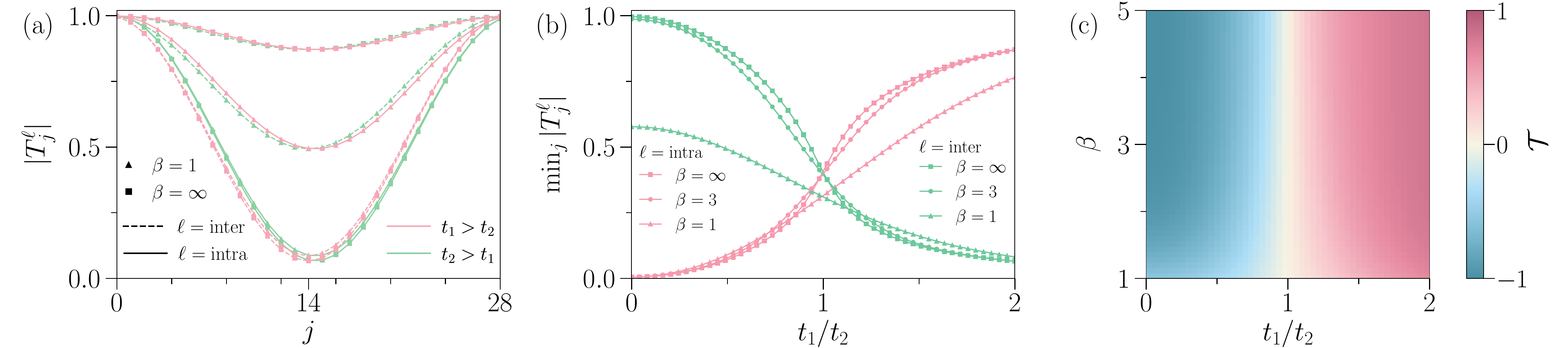}  
    \caption{(a): The magnitude of the expectation value of the local twist operator, $T_j^{\ell}$, where $\ell$ corresponds to either intra or inter, as a function of cell position $j$ for three values of inverse temperature $\beta$, and for the two phases where $t_1=2, t_2=1$ colored in green, and $t_1=1, t_2=2$ colored in pink.
    $\ell=$intra is depicted with a solid line, and $\ell=$inter is depicted with a dashed line. 
    Square markers correspond to $\beta=\infty$ (zero temperature), and circular markers to $\beta=1$. 
     (b): The minimum value of the magnitude of the expectation values of the intra (pink) and inter (green) cell operators as a function of $t_1/t_2$ where $t_2=2$, for $N=19$. 
     The square, circular, and triangular markers correspond to $\beta=\infty$, $\beta=3$, and $\beta=1$ respectively. 
     (c): $\mathcal{T}$ as a function of $\beta$ and $t_1/t_2$, where $t_2=2$, and $N=19$.
     The unit-cell centers are at positions $x_j=j$, with $j=0,...,N-1$ ($N$ odd), with orbital positions $x_j^A=x_j-\delta x$, $x_j^B=x_j+\delta x$ with $\delta x=1/4$.}
    \label{fig:panel}
\end{figure*}

\subsection{The phase of the local operators}

Consider the total phase of the intracell operators, 
\begin{align}
\varphi^{\rm intra} = \sum_j \Im \ln \langle T_j^{\rm intra}\rangle 
= \sum_j \arg\!\big(\langle T_j^{\rm intra} \rangle\big),
\end{align}
for positive hopping amplitudes $t_1>0$ and $t_2>0$. 
To analyze this quantity, it is convenient to place the inversion center of the SSH chain at  position $x_0=0$. 
For each site $j$ there exists an inversion-symmetric partner $\bar{j}$ with  $x_{\bar{j}}=-x_j$, such that
\begin{align}
\langle T_{\bar{j}}^{\rm intra}\rangle 
= \langle T_j^{\rm intra}\rangle^*,
\label{eq:inversionsymTi}
\end{align}
which implies that the arguments of the expectation values for inversion symmetric pairs are equal, but with opposite sign:
\begin{align}
\arg\!\big(\langle T_{\bar{j}}^{\rm intra}\rangle\big) 
= -\,\arg\!\big(\langle T_j^{\rm intra}\rangle\big).
\end{align}
For an even number of unit cells $N$, the phases therefore cancel pairwise, such that the total phase is $\varphi^{\rm intra}=0$. 
For odd $N$, the central site, $x_0$ lies at the inversion center, and Eq.~\eqref{eq:inversionsymTi} ensures that its expectation value is real, which means that the phase contribution from $x_0$ only be $0$ or $\pi$. 
In the strict trivial limit, $t_2=0$,  Eq.~\eqref{eq:Tintra-triv} is positive, and $ \varphi_0^{\rm intra}=0$.
In fact, $ \varphi^{\rm intra}$ is always zero regardless of the relative size of the inter and intracell hopping parameters. 
Changing the phase of the real expectation value would require the expectation value itself to change sign, and thus to pass through zero.
As shown in Appendix~\ref{appendix:phase}, the modulus of the intracell operator remains strictly positive for all values of $t_1/t_2$, and at any temperature, preventing such a sign change.
This means that $\varphi^{\rm intra}=\varphi_{\rm{EGP}}$ only in the trivial phase.

A similar argument applies to the intercell operators, which act on the bonds between unit cells. 
With periodic boundary conditions there are $N$ such bonds, where $T_j^{\rm inter}$ acts on the bond connecting sites $B_j$ and $A_{j+1}$ in unit cells $j=0,\dots,N-2$, and the bond between the first and last unit cell connects sites $B_{N-1}$ and $A_0$.
The average position of $T_j^{\rm inter}$ is the bond center $x_j^{\rm bond}=(x_j+x_{j+1})/2$, which is mapped by inversion as $x_j^{\rm bond}\mapsto -x_j^{\rm bond}$.
As a consequence, most intercell operators appear in inversion-related pairs, whose phases cancel in the total sum. 
The only exceptions are the bonds that are mapped onto themselves. 
For odd $N$, there is exactly one such bond, located at the chain center with index $j=(N-1)/2$. 
The corresponding phase $\varphi^{\rm{inter}}_{(N-1)/2}=\pi$ for any positive values of the hopping parameters (see appendix~\ref{appendix:phase}).
For even $N$, there are two bonds that map to themselves: the central bond at $j=N/2-1$ and the bond connecting the unit cells at the two ends of the chain, at $j=N-1$. 
Here the corresponding phases are $\varphi^{\rm{inter}}_{N/2-1}=\pi$, and $\varphi^{\rm{inter}}_{N-1}=0$ (see appendix~\ref{appendix:phase}).
The total phase is therefore always equal to the phase of the operator at the central bond, and $\varphi^{\rm{inter}}=\varphi_{\rm{EGP}}$ only in the topological phase, $t_2>t_1$.

The local twist operators $T_j^{\mathrm{intra}}$ and $T_j^{\mathrm{inter}}$ capture the topology of the SSH chain, but their significance depends on the phase: $T_j^{\mathrm{intra}}$ characterizes the trivial regime, while $T_j^{\mathrm{inter}}$ characterizes the topological one.
To use these operators as practical indicators of topology, one must determine which is relevant without assuming prior knowledge of the phase.
This is achieved by comparing the magnitudes of the two operators.

\subsection{The magnitude of the local twist operators}

The modulus of the expectation values of the inter and intracell operators is essentially independent of the system size $N$. 
Both $T_j^{\rm intra}$ and $T_j^{\rm inter}$ contain diagonal factors of the form $\e^{2\pi i x_j/N}$, so the functional dependence of $|\langle T_j\rangle|$ on $j$ is controlled by the 
scaled coordinate $x_j/N$ rather than by $N$ itself. 
As a result,  the modulus is stable as $N$ increases, with the minimum located in the middle of the chain and the values enhanced near the boundaries, as depicted in Fig.~\ref{fig:panel} (a).
Fig.~\ref{fig:panel}(b) shows that the modulus of $T_j^{\rm intra}$ exceeds that of $T_j^{\rm inter}$ in the trivial phase, whereas the opposite holds in the topological phase. 
This behavior reflects the localization of the Wannier centers on atomic sites in the trivial phase and on bonds in the topological phase. 
Consequently, the phase of a given chain can be characterized by measurements of two local expectation values: one of the intracell operator and one of the intercell operator, both taken at the center of the chain.
We define the quantity
\begin{align}
\mathcal{T}
= |\langle T_{\mathrm{c}}^{\mathrm{intra}}\rangle|
 - |\langle T_{\mathrm{c}}^{\mathrm{inter}}\rangle|,
\end{align}
where the position label $\mathrm{c}$ denotes the unit cell at the center of the chain.
The quantity $\mathcal{T}$ is positive in the trivial phase and negative in the topological phase, crossing zero at the phase transition, as shown in Fig.~\ref{fig:panel}(c).
Thus, $\mathcal{T}$ provides a local indicator of the topological phase, requiring only measurements of two local operators.

\subsection{Inversion-symmetric chain with next-nearest-neighbor coupling}

Adding an inversion-symmetry-preserving next-nearest-neighbor hopping $t_3$ extends the SSH model’s topological phase diagram at zero temperature to three distinct regimes.  
Inversion symmetry constrains the Wannier centers to be located either at the centers of the unit cells or halfway between them.  
Depending on the relative magnitudes of $t_1$, $t_2$, and $t_3$, the Wannier center localizes at site $j$ when $t_1 > t_2 + t_3$; between two unit cells, at $j + 1/2$, when $t_1 < t_2 + t_3$ and $t_2 > t_3$; and at site $j + 1$ when $t_1 < t_2 + t_3$ and $t_2 < t_3$. %
For $t_3 > t_2$, the ensemble geometric phase does not distinguish the two topological phases with Wannier centers at $j+1/2$ and $j+1$, and the single-site intra and intercell operators considered so far also fail to capture the shift of Wannier weight between unit cells.
The Wannier function becomes delocalized between $j + 1/2$ and $j + 1$ at the transition where $t_2=t_3$.  
In this regime, a single-site intracell operator no longer captures the correct spatial structure, since the relevant weight is distributed over two neighboring unit cells.  
To account for this, we replace the single-site operator $T_j^{\rm intra}$ with a two-site intracell operator that acts on both sublattices in adjacent unit cells,
\begin{align}
T_j^{\rm 2\,intra}
&=\exp\!\left\lbrace\i\,\delta k\!\sum_{\ell=j}^{j+1}\big[x_\ell n_\ell+\delta x(n_\ell^{B}-n_\ell^{A})\big]\right\rbrace,
\label{eq:T2intra}
\end{align}
which better resolves the change in localization across the phase transition. 
We also introduce an operator acting on the same bond as the next-nearest-neighbor hopping term, $t_3$, in Eq.~\eqref{eq:H_SSH_nnn},
\begin{align}
T_j^{\rm nnn}
&=\exp\!\left\lbrace\i\,\delta k\big[x_j n_j^{A}+x_{j+1} n_{j+1}^{B}+\delta x(n_{j+1}^{B}-n_j^{A})\big]\right\rbrace.
\label{eq:Tnnn}
\end{align}
which, like $t_3$, is nonlocal and connects site $A$ in unit cell $j$ to site $B$ in unit cell $j+1$; see Fig.~\ref{fig:operators}.  
To determine whether a given phase is trivial or topological, we define the quantity
\begin{align}
\mathcal{T}_{3}
= |\langle T_{\mathrm{c}}^{\mathrm{2\,intra}}\rangle|
- M,
\end{align}
where \( M = \max\bigl(|\langle T_{\mathrm{c}}^{\mathrm{inter}}\rangle|,\; |\langle T_{\mathrm{c}}^{\mathrm{nnn}}\rangle|\bigr) \), and the unit cell at $j=\rm{c}$ corresponds to the cell in the center of the chain.
At zero temperature, $\mathcal{T}_{3}<0$ signals a topological phase as shown in Fig.~\ref{fig:operators}(a).
To distinguish between the two topological phases, we define 
\begin{align}
\mathcal{I}
= |\langle T_{\mathrm{c}}^{\mathrm{inter}}\rangle|
-|\langle T_{\mathrm{c}}^{\mathrm{nnn}}\rangle|,
\end{align}
which is positive when $t_2>t_3$, and negative when $t_2<t_3$ as shown in Fig.~\ref{fig:operators}(b).
The superimposed line in Fig.~\ref{fig:operators}(a) shows the locus where $\mathcal{T}_{3}=0$.
This line coincides with the phase transition between 
the trivial and topological phases, except in a region close to the triple phase transition at $t_{3}/t_{1}=t_{2}/t_{1}$.
In this region the curve develops a cusp, reflecting that although 
$|\langle T_{\mathrm{c}}^{\mathrm{2\,intra}}\rangle|$ decays rapidly along the line $1 - t_{2}/t_{1}$,
it still decays more slowly than both 
$|\langle T_{\mathrm{c}}^{\mathrm{inter}}\rangle|$ and 
$|\langle T_{\mathrm{c}}^{\mathrm{nnn}}\rangle|$ near the triple phase transition.
To capture the phase in this region, it is therefore necessary to additionally consider the value of $\mathcal{I}$, which remains close to zero in the trivial region near the transition between the three phases.

Panels (c) and (d) in Fig.~\ref{fig:operators} display $\mathcal{T}_{3}$ and $\mathcal{I}$ evaluated at $t_{1}=2$ and $t_{2}/t_{1}=0.75$, as functions of inverse temperature $\beta$ and relative hopping $t_{3}/t_{1}$. The results show that the region around the zero-temperature phase transition becomes progressively smeared out as temperature increases.
The temperature dependence of $\mathcal{T}_{3}$ and $\mathcal{I}$ highlights how thermal fluctuations weaken the distinction between the competing hopping processes and broaden the localization of the Wannier centers.

While the local twist operators are not topological invariants, their relative magnitudes offer a local signature of the  topological and trivial phases.
When the purity gap closes, the local twist operators remain well defined as expectation values, but they cease to provide a meaningful diagnostic of mixed-state topology.
In the maximally mixed limit \(\beta\to 0\), the expectation values of the local twist operators become independent of the Hamiltonian, and the magnitudes of the central intracell and intercell twist operators vanish with system size, so they no longer carry information about the topological phase.

\begin{figure}[t!]
    \centering
    \includegraphics[width=\columnwidth]{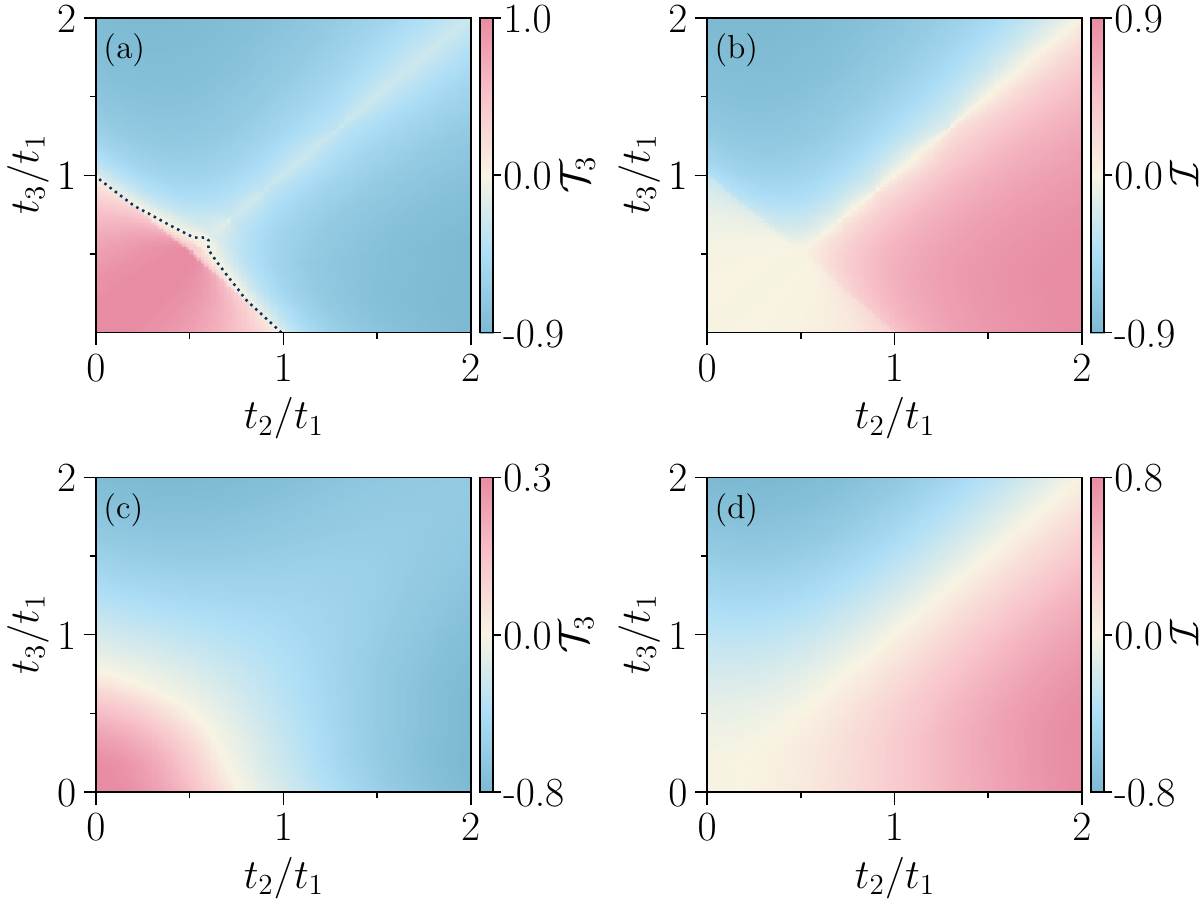}  
    \caption{(a): $\mathcal{T}_{3}$ as a function of the ratios $t_3/t_1$ and $t_2/t_1$,the hopping parameters of the next-nearest-neighbor SSH model, at zero temperature. The dotted line denotes $\mathcal{T}_{3}=0$. (b): $\mathcal{I}$, as a function of $t_3/t_1$ and $t_2/t_1$ at zero temperature, where the difference in sign distinguishes between two topological phases. (c): $\mathcal{T}_{3}$ as a function of inverse temperature $\beta$ and $t_2/t_1$. (d): $\mathcal{I}$ as a function of inverse temperature $\beta$ and $t_2/t_1$. $t_1=2$ in all panels, and $t_3=1.5$ in panels (c) and (d). The number of unit cells is $N=32$ in all four panels. The unit-cell centers are defined as $x_j=j+1/2$, with $j=0,...,N-1$ ($N$ even), with orbital positions $x_j^A=x_j-\delta x$, $x_j^B=x_j+\delta x$ with $\delta x=1/4$.}
    \label{fig:NNN}
\end{figure}

The local twist operators are diagonal in the occupation basis and are therefore directly accessible in cold-atom quantum gas microscope experiments with single-site-resolved readout~\cite{Cheuk2015,Haller2015,Parsons2015,Gross2017}.
In each projective snapshot $s$, the measured occupations $n_j^{(s)}\in\{0,1\}$ determine a corresponding $c$-number value $T_{j,\ell}^{(s)}$ for $\ell\in\{\mathrm{intra},\mathrm{inter},\mathrm{nnn}\}$.
The expectation value is then obtained by averaging over many shots,
\begin{align}
\langle T_j^{\ell}\rangle \approx \frac{1}{N_{\rm shots}}\sum_s T_{j,\ell}^{(s)}.
\end{align}
By contrast, the global twist operator defining the ensemble geometric phase requires reliable readout over the full chain and suffers from reduced contrast at nonzero temperature and large system size.

\section{The local chiral marker}

The local chiral marker is a real space topological invariant that is equal to the chiral winding number for translation-invariant structure and characterizes the topology of states with a chiral constraint.
It belongs to a class of local topological markers, including the local Chern marker~\cite{Bianco2011}, the local chiral, and the local Chern-Simons marker~\cite{Hannukainen2022}, formulated in terms of the one-particle density matrix.
The local chiral marker applies to chiral-symmetric states in symmetry classes and spatial dimensions where the bulk classification is $\mathbb{Z}$-valued.

The Bogoliubov-de-Gennes one-particle density matrix of a many-body state $|\Psi\rangle$ is
\begin{align}
\varrho =\begin{pmatrix}
\tilde{\varrho} &\kappa\\
\kappa^\dagger&1-\tilde{\varrho^*}
\end{pmatrix},
\end{align}
where $\tilde{\varrho}=\langle \Psi|c_i^\dagger c_j|\Psi\rangle$, and $\kappa=\langle \Psi|c_i^\dagger c_j^\dagger|\Psi\rangle$, and $c_i$, $c_j^\dagger$ are fermionic annihilation and creation operators respectively~\cite{Bera2015,Bera2017,Kells2018}.
For number-conserving Gaussian states, such as the thermal states of the SSH Hamiltonians considered in this work, the anomalous correlator vanishes identically, $\kappa=0$, and $\varrho$ reduces to the normal one-particle density matrix $\tilde{\varrho}$.
The spectrum of $\varrho$ takes values in the interval $n_\alpha\in [0,1]$, and are interpreted as the occupations of the natural orbitals $|n_\alpha\rangle$, the diagonalized eigenbasis of $\varrho$~\cite{Penrose1956,Koch2001,Bera2015,Bera2017,Lezama2017}. 
For a translation-invariant pure Gaussian state, the one-particle density matrix acts as a projector onto the occupied states in momentum space.  
The image of the projector defines a subspace of the Hilbert space at each momentum in the Brillouin zone,  and the collection of these subspaces across all momenta forms a vector 
bundle. 
The topology of this bundle encodes the topological phase of the state. 
The projector commutes with the unitary symmetry operators, yielding a block-diagonal structure.
Each block defines a vector bundle, further constrained by antiunitary or chiral symmetries, and the  resulting symmetry-constrained bundles are classified within the Altland--Zirnbauer scheme.
In particle-number-conserving symmetry classes and in the absence of pairing correlations, the topological classification takes as input the correlation matrix
\begin{align}
\tilde{\varrho}_{ij}=\langle \Psi|c_i^\dagger c_j|\Psi\rangle.
\end{align}

Two projectors are topologically equivalent if they can be continuously deformed into one another without closing the spectral gap, while preserving the protecting symmetries and the exponential locality of their matrix elements.
If the one-particle density matrix is not a projector, it no longer define a vector bundle over the Brillouin zone.
However, if the one-particle density matrix has a spectral gap, it can be adiabatically flattened into a projector while retaining the topological properties of the original state.
This approach has been employed to classify interacting systems with a gapped one-particle density matrix using the local chiral marker~\cite{Hannukainen2024}.

We use the local chiral marker to characterize the topology of the SSH chain at nonzero temperature, by using the topological classification of mixed Gaussian states with a purity gap.
A mixed fermionic Gaussian state $\rho~\sim \e^{-G}$ is fully determined by its two-point correlations $f(G)$, Eq.~\eqref{eq:C-from-G}, which reduces to a projector onto the filled energy bands in the zero-temperature limit.
At nonzero temperature $f(G)$ is no longer a projector, but if the state has a purity gap, its spectrum remains gapped and can be adiabatically flattened into a projector, Eq.~\eqref{eq:P-from-fG}, which carries the same topology as the original mixed state.
Consequently, the topology of the mixed state can be characterized by evaluating the local chiral marker on the projector in Eq.~\eqref{eq:P-from-fG}.

\begin{figure}[t!]
    \centering
    \includegraphics[width=1\linewidth]{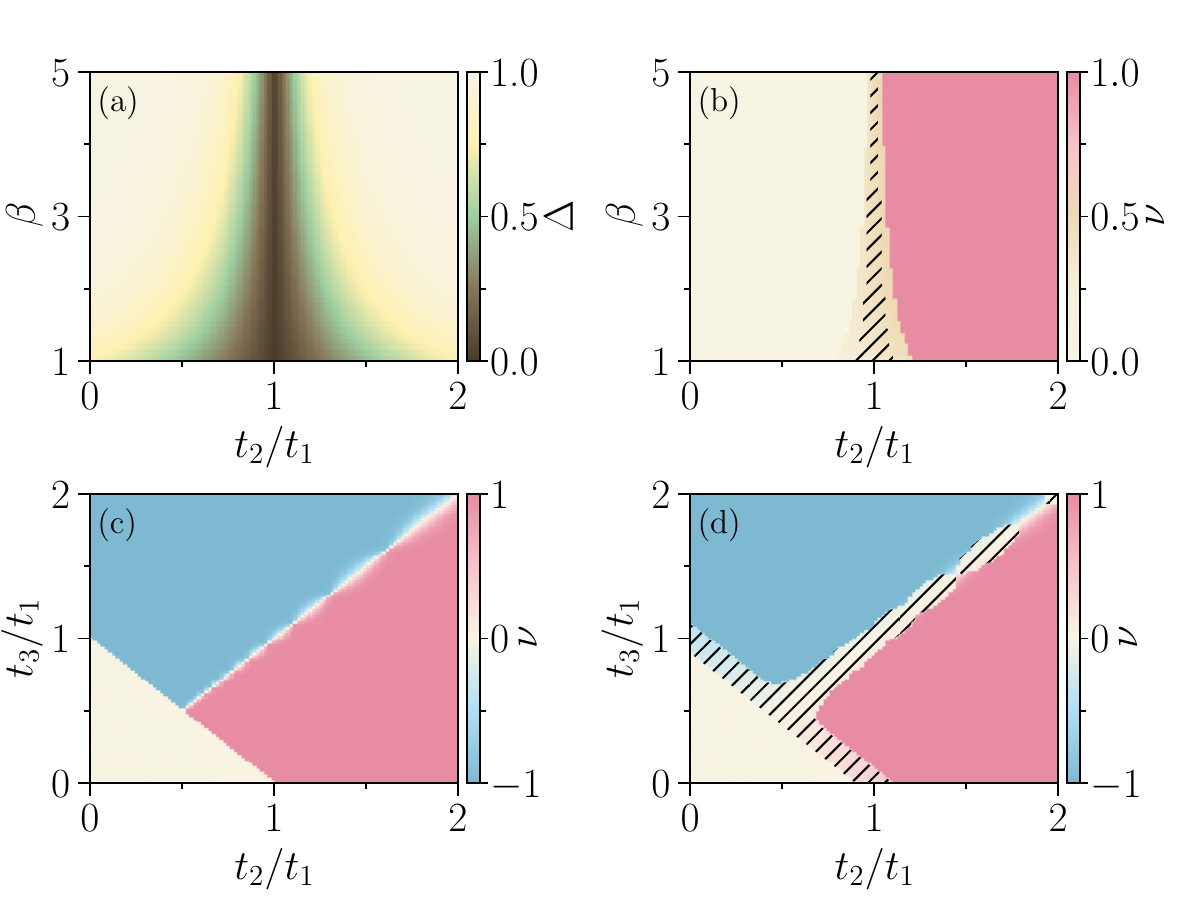}  
    \caption{(a): The  gap $\Delta$ of the spectrum of the correlation matrix $f(H_{\rm{SSH}})$ as a function of inverse temperature $\beta$ and the ratio of the intra and inter hopping parameters $t_2/t_1$.  (b): The chiral marker $\nu$ as a function of $\beta$ and the ratio $t_2/t1$ for the SSH model with only nearest-neighbor hopping. The stripes denote the region where $\Delta<0.1$ (c): The chiral marker $\nu$ at zero temperature as a function of $t_3/t_1$ and $t_2/t_1$ for the SSH model including next-nearest-neighbor hopping. (d): The chiral marker $\nu$ for $\beta=1$ as a function of $t_3/t_1$ and $t_2/t_1$.  The stripes denote the region where $\Delta<0.1$. The inter cell hopping is $t_1=2$, and the number of unit cells is $N=80$ in all four panels.}
    \label{fig:marker}
\end{figure}

The local chiral marker is defined as~\cite{Hannukainen2022}
\begin{align}
\nu(x)=-2\sum_{\alpha}\langle \alpha, x|P S X P| \alpha, x\rangle,
\end{align}
where $P$ is the band flattened correlation matrix, Eq.~\eqref{eq:P-from-fG}, $S=-\sigma_z\otimes \id_N$ is the chiral constraint of the SSH Hamiltonian, $X$ is the position operator, and the trace over $\alpha$ runs over the two sublattice degrees of freedom.
All marker calculations in this work are performed with periodic boundary conditions.
For translation-invariant structures, the local chiral marker is spatially uniform.
Thus, in Fig.~\ref{fig:marker}(b)–(d), the marker takes the same value at every bulk position, and no spatial averaging is required.
The SSH model with only nearest-neighbor hopping exhibits two insulating phases, separated by a topological phase transition at \(t_1 = t_2\) at zero temperature.
Fig.~\ref{fig:marker}(a) shows the corresponding purity gap \(\Delta\) of the correlation matrix \(f(H_{\mathrm{SSH}})\) as a function of inverse temperature \(\beta\) and hopping ratio \(t_2/t_1\). 
The purity gap is equal to one away from the phase transition for the pure zero temperature state.
As temperature increases, the Fermi--Dirac occupations become smeared, mixing states above and below the chemical potential which reduces \(\Delta\) even away from the transition and broadens the region where it becomes small. 
This behavior marks a gradual crossover from pure-state topology to a thermally mixed regime with increasing temperature.
The corresponding chiral marker \(\nu\), shown in Fig.~\ref{fig:marker}(b), takes quantized values in the gapped regions, $\nu = 1$ in the topological phase and $\nu = 0$ in the trivial one.
The hatched area indicates where the purity gap falls below the numerical threshold \(\Delta < 0.1\) used for band flattening.
When next-nearest-neighbor hopping $t_3$ is introduced, the zero temperature phase diagram expands to include a third distinct phase, as shown in Fig.~\ref{fig:marker}(c), where \(\nu = 0\) and \(\pm1\) characterize the three phases.
At nonzero temperature with \(\beta = 1\), shown in Fig.~\ref{fig:marker}(d), these features persist but become less sharply resolved.
This reflects the reduction of the purity gap with increasing temperature, which lowers the numerical contrast of the marker near the underlying zero-temperature transition lines while leaving the mixed-state topology unchanged for any finite \(\beta\) as long as the gap remains open.
All results are obtained for intercell coupling \(t_1 = 2\) and a chain of \(N = 80\) unit cells.

\section{Discussion}

We have compared three complementary approaches for diagnosing topology in mixed Gaussian states at nonzero temperature, focusing on the SSH chain and its inversion-symmetric extension.
These approaches are the ensemble geometric phase, local twist operators, and the local chiral marker.
At zero temperature, all three methods reproduce the same topological information for the nearest-neighbor SSH model. 
For the SSH model with nearest-neighbor hopping, the ensemble geometric phase coincides with the quantized Zak phase of the lower band, distinguishing between a topologically trivial and nontrivial phase.
The local twist operators—intracell and intercell—encode the same phase distinction: the former reflects the dominant intracell bond in the trivial phase,  while the latter captures the intercell bond characteristic of the topological phase.
The relative magnitude of the two operators reverses across the phase transition, with the intracell term dominant in the trivial phase, and the intercell term dominant in the topological phase.
Comparing their magnitudes at the center of the chain therefore provides a direct and local diagnostic of the phase.
The local chiral marker coincides with the winding number in translation-invariant models.
With next-nearest-neighbor hopping, the local twist operators, and the chiral marker resolve the three topological sectors, while the ensemble geometric phase, being a phase, remains quantized modulo \(2\pi\), and does not distinguish between the two different topologically nontrivial phases.

At nonzero temperature, the three methods remain consistent in identifying the phase transition, but they capture the topology through different quantities and differ in practicality.
In the thermal SSH family considered here, temperature does not act as an independent tuning parameter that changes the topological phase at finite $\beta$; instead it controls the purity gap and the visibility of the diagnostics.
The ensemble geometric phase is well-defined for any \(T>0\) with a nonzero purity gap, yet we showed how the modulus \(|\langle T\rangle|\) decays exponentially with system size, which limits its applicability in large systems.
The relative magnitudes of the intra-, inter-, and next-nearest-neighbor twist operators decrease with increasing temperature,  but they continue to distinguish the phases and coincide at the corresponding phase boundaries.
This temperature dependence demonstrates that the zero-temperature criterion for topological characterization---based on comparing the magnitudes of the local twist operator---remains valid in the thermal regime.
We formulated the local chiral marker for mixed Gaussian states with a purity gap by using the band-flattened single-particle correlation matrix.
Evaluating the chiral marker with respect to this projector produces a real-space invariant that matches the zero-temperature winding number and remains well behaved at finite temperature as long as a purity gap is present.

Each of the three methods serves a distinct purpose.
The ensemble geometric phase offers a direct mixed-state generalization of geometric polarization but suffers from a vanishing modulus in the thermodynamic limit.
The local twist operators provide a minimal, experimentally accessible diagnostic requiring only local measurements and naturally extend to distinguish the three topological phases in the extended SSH model, serving as local indicators of topology rather than topological invariants.
Their diagonal form in the occupation basis makes them particularly natural for single-site-resolved measurements in cold-atom quantum gas microscope platforms.
The chiral marker provides a real-space topological invariant that applies beyond translation-invariant settings and aligns with the projector-based classification of Gaussian mixed states.

All three approaches rest on the mixed-state classification of Gaussian systems possessing a purity gap, which guarantees adiabatic continuity to a pure state and allows for the band flattening of the correlation matrix.
When the purity gap closes---either around the zero-temperature topological transition or at sufficiently high temperature---the mixed-state classification ceases to apply.
Within the gapped regime, however, the three methods remain mutually consistent and provide complementary means of characterizing topology at nonzero temperature.
The three approaches remain valid at weak disorder whenever the single-particle spectrum retains a finite spectral or mobility gap at zero temperature.
In this regime, the purity gap of the Gaussian state inherits the spectral or mobility gap, ensuring adiabatic continuity to the pure-state limit.
We expect weak interactions to leave this qualitative behavior intact, with all three diagnostics continuing to agree as long as the purity gap remains.

\section{Acknowledgements}
We thank Frank Schindler,  Miguel Martínez, and Tom Sheppard for valuable discussions.
This work was supported by EPSRC (Grant Nos. EP/Y01510X/1 and EP/V062654/1) a Simons Investigator Award (Grant No. 511029).

\appendix

\section{Highest-order correction to $|\langle T\rangle|$ }

\label{app:perturbative_expansion}

The expectation value of the many-body translation operator $T$ is given by
\begin{equation}
\langle T \rangle = \frac{1}{Z}\,\det\!\left[\sigma_0 + M_{\mathrm{T}}\right],
\label{eq:appendix-T_expvalue}
\end{equation}
defined in Eq.~\eqref{eq:T_expvalue}
In the main text we expanded the expectation value perturbatively in the off-diagonal transition matrices describing inter band processes, and showed that the zeroth-order term grows exponentially with system size.
To evaluate the highest-order correction,
\begin{align}
    \Delta(N) = \Tr(G V^{(2)}) - \tfrac{1}{2}\Tr[(G V^{(1)})^2],
\end{align}
appearing in the expansion of $\ln \det[\sigma_0 + M_{\mathrm{T}}]$, Eq.~\eqref{eq:V_expansion}, it is convenient (following the approach of Ref.~\cite{Bardyn2018}) to introduce the projectors
\begin{align}
    P^{\pm} = \tfrac{1}{2}(\sigma_0 \pm \sigma_3),
\end{align}
which project onto the upper ($+$) and lower ($-$) bands, respectively. 
The matrices $\sigma_i$ ($i=1,2,3$) denote the Pauli matrices, and $\sigma_0$ is the identity matrix in band space.
The Green's function can then be written as
\begin{align}
    G = P^{+} +(W^{-}_{2\pi,0})^{-1} P^{-},
\end{align}
where the diagonal operators $W^\pm_{k_1, k_2} = \exp[ \int_{k_1}^{k_2} dk \, (\i A_k^{\pm} \pm \delta k^{-1} \beta \varepsilon_k)] \sigma_0$, acts within a single band, with $A_k^\pm = \mathcal{A}_k^0 \pm \mathcal{A}_k^3 =\i \langle u_k^\pm | \partial_k u_k^\pm \rangle$
The off-diagonal components of the perturbation are given by
\begin{align}
    V_k^{+-} = (\mathcal{A}_k^1 - \i \mathcal{A}_k^2)\,\sigma^{+}, \qquad
    V_k^{-+} = (\mathcal{A}_k^1 + \i \mathcal{A}_k^2)\,\sigma^{-},
\end{align}
with $\sigma^{\pm} = (\sigma_1 \pm \i\sigma_2)/2$.
In the large-$N$ limit, the first- and second-order terms in the expansion take the form
\begin{align}
    V^{(1)} &= \int_{0}^{2\pi} \! dk \, \big(
        W^{+}_{2\pi,k} V_k^{+-} W^{-}_{k,0}
        + W^{-}_{2\pi,k} V_k^{-+} W^{+}_{k,0}
    \big), \\
    V^{(2)} &= \int_{0}^{2\pi} \! dk \int_{0}^{k} \! dk' \, \big(
        W^{+}_{2\pi,k} V_k^{+-} W^{-}_{k,k'} V_{k'}^{-+} W^{+}_{k',0} \nonumber\\
        &\hspace{2cm}+ W^{-}_{2\pi,k} V_k^{-+} W^{+}_{k,k'} V_{k'}^{+-} W^{-}_{k',0},
    \big).
\end{align}
such that, after using the cyclic property of the trace together with the identities,  $(W^{\pm}_{k_1 k_2})^{-1}=W^{\pm}_{k_2 ,k_1}$, and $W^{\pm}_{k_1 ,k_2}W^{\pm}_{k_2 ,k_3}W^{\pm}_{k_1 ,k_3}$, the correction becomes
\begin{align}
    \Delta(N) &=\Tr \int_0^{2\pi} dk \int_0^k dk' \Big[\nonumber \\
     & +(W^{-}_{k, k'})^{-1} V^{-+}_k W^{+}_{k, k'} V^{+-}_{k'}\nonumber \\
    &+ W^{+}_{2\pi, 0} (W^{+}_{k, k'})^{-1} V^{+-}_k W^{-}_{k, k'} V^{-+}_{k'} \nonumber \\
    & - (W^{-}_{k, k'})^{-1} V^{-+}_k W^{+}_{2\pi, 0} W^{+}_{k, k'} V^{+-}_{k'}  \nonumber\\
    &-W^{+}_{2\pi, 0} (W^{+}_{k, k'})^{-1} V^{+-}_k W^{-}_{k, k'} V^{-+}_{k'} \Big].
        \label{eq:2ndorder_correction}
\end{align}
Only the first line in Eq.~\eqref{eq:2ndorder_correction} contributes in the large-$N$ limit.  
The second and fourth lines are equal in magnitude but opposite in sign, while the third line contains the factor $W^{+}_{2\pi, 0}\propto \e^{- N\beta \bar{\varepsilon}}$, with $\bar{\varepsilon} = \int_0^{2\pi} \! dk\, \varepsilon_k$, which is exponentially suppressed.  

The quantity of interest is the modulus $|\langle T \rangle|$, whose leading correction arises from the real part of $\Delta(N)$.  
Focusing on the real contribution to Eq.~\eqref{eq:2ndorder_correction} gives
\begin{align}
&\mathrm{Re}\,\Delta(N)
= \int_0^{2\pi} {\rm d}k \int_0^k dk'\;
\e^{2 \int_{k'}^k {\rm d}q\,(- \delta k^{-1}\beta \varepsilon_q)}\nonumber\\
&\times (\mathcal{A}_k^1 - \i \mathcal{A}_k^2)
(\mathcal{A}_{k'}^1 + \i \mathcal{A}_{k'}^2) \cos\!\left(2\!\int_{k'}^k {\rm d}q\, \mathcal{A}_q^3 \right). 
\label{eq:ReDelta_def}
\end{align}
The factor
\begin{equation}
\e^{2\!\int_{k'}^{k}\! {\rm d}q\,(-\delta k^{-1}\beta\,\varepsilon_q)}
\simeq \e^{-B_k \xi}, \qquad  
B_k = \tfrac{2\beta\,\varepsilon_k}{\delta k},
\end{equation}
in Eq.~\eqref{eq:ReDelta_def} strongly suppresses contributions from large momentum separations $\xi = k - k'$. 
At large $N$ (small $\delta k$), the integral is dominated by small $\xi$,  where $\mathcal{A}_k^i$ and $\varepsilon_k$ vary smoothly and can be expanded as
\begin{align}
   & \mathcal{A}_{k'}^{i} = \mathcal{A}_{k}^{i} - \xi\,\partial_k \mathcal{A}_k^{i} + \mathcal{O}(\xi^2), \\
    &\int_{k'}^k dq\,\beta \varepsilon_q = \xi\,\beta \varepsilon_k+ \mathcal{O}(\xi^2),\label{eq:betaexpansion}
\end{align}
The oscillatory term 
$\cos\!\left(2\!\int_{k'}^{k}\! dq\,\mathcal{A}_q^3\right)\simeq \cos(2\,\mathcal{A}_k^3\,\xi)$ remains bounded by unity and varies slowly compared to the exponential decay.
In the small $\xi$ expansion the real contribution reads 
\begin{align}
\mathrm{Re}\,\Delta(N)
&\simeq \int_0^{2\pi}\! dk\,\mathcal{A}_{k,\perp}^2
\int_0^{k}\! d\xi\, \e^{-B_k \xi}\,
\cos\!\big(2\,\mathcal{A}_k^3\,\xi\big),
\label{eq:ReDelta_LO_short}
\end{align}
where $\mathcal{A}_{k,\perp}^2 \equiv (\mathcal{A}_k^1)^2 + (\mathcal{A}_k^2)^2$, and $B_k \equiv \tfrac{2\,\beta\,\varepsilon_k}{\delta k}$ (more generally, $B_k \equiv 2\,\delta k^{-1}\!\int_{k-\xi}^{k}\! {\rm d}q\,\beta_q$ evaluated at $k$). 
For $B_k  k\gg 1$, the exponential decay suppresses contributions near the Brillouin-zone boundary, allowing the upper limit of the $\xi$-integral to be extended to infinity with exponentially small error, such that
 \begin{align}
  \mathrm{Re}\,\Delta(N) &\approx \int_0^{2\pi}\! dk\,\big[(\mathcal{A}_k^1)^2+(\mathcal{A}_k^2)^2\big]\, \frac{B_k}{B_k^2+\big(2\,\mathcal{A}_k^3\big)^2}. \label{eq:ReDelta_closed_short}
 \end{align}
 Since
\begin{align}
B_k \;\ge\; \frac{2\,\Delta\beta}{\delta k}
\;=\; \frac{N\,\Delta\beta}{\pi}\,,
\end{align}
where $\Delta\beta = \min_k (\beta\,\varepsilon_k) > 0$ is the minimal value of $\beta\,\varepsilon_k$ on the Brillouin zone, ensuring exponential damping for the smallest gap, and using
\begin{align}
\dfrac{B_k}{B_k^2+\omega_k^2} \le \dfrac{1}{B_k},
\end{align}
the correction is
\begin{align}
\mathrm{Re}\,\Delta(N)
&\le \int_0^{2\pi}\! {\rm d}k\,\frac{(\mathcal{A}_k^1)^2+(\mathcal{A}_k^2)^2}{B_k}\nonumber\\
&\le\; \frac{\pi}{N\,\Delta\beta}
\int_0^{2\pi}\! {\rm d}k\,\big[(\mathcal{A}_k^1)^2+(\mathcal{A}_k^2)^2\big].
\end{align}
So $\mathrm{Re}\,\Delta(N)=\mathcal{O}(1/N)$ for smooth, $k$-bounded
$\mathcal{A}_k^i$ and a nonzero gap ($\Delta\beta>0$).

The modulus of the numerator in Eq.~\eqref{eq:T_expvalue} therefore behaves as
\begin{align}
|\det(\sigma_0+M_{\mathrm{T}})|
\simeq \e^{N\beta\bar{\varepsilon}}\!\left(1 + \mathcal{O}(N^{-1})\right),
\end{align}
for large \(N\).

\section{The phase of the twist operators at inversion-symmetric points }
\label{appendix:phase}

At the inversion center, the intracell operator is defined as 
\begin{align}
T^{\mathrm{intra}}_0
&=\exp(\i\,\delta k\,X^{\mathrm{intra}}_0),
\end{align}
where $\delta k=2\pi/L$ for a chain of length $L$, and 
\begin{align}
X^{\mathrm{intra}}_0
&=x_0^{A} n_0^{A}+x_0^{B} n_0^{B},\qquad
x_0^{B}=-x_0^{A}.
\end{align}
being the position operator.
The intracell operator $T^{\mathrm{intra}}_0$ is supported on the  spatial region that inversion leaves invariant. 
Since the position operator is inversion-odd,  $\mathcal I X^{\mathrm{intra}}_0 \mathcal I^{-1}=-X^{\mathrm{intra}}_0$, the intracell operator transforms as 
\begin{align}
\mathcal I\,T^{\mathrm{intra}}_0\,\mathcal I^{-1}
&=(T^{\mathrm{intra}}_0)^\dagger,\\
\mathcal I\,\e^{-\beta H_{\mathrm{SSH}}}\,\mathcal I^{-1}
&=\e^{-\beta H_{\mathrm{SSH}}},
\end{align}
meaning that inversion reverses the phase of $T^{\mathrm{intra}}_0$ but does not move it away from the inversion center.  
Because the thermal density matrix $\e^{-\beta H_{\mathrm{SSH}}}$ commutes with $\mathcal I$, the expectation value of any operator and that of its inversion partner are equal, 
\begin{align}
\langle \mathcal I O \mathcal I^{-1}\rangle
= \mathrm{Tr}\!\left(\e^{-\beta H_{\mathrm{SSH}}}\mathcal I O \mathcal I^{-1}\right)
= \mathrm{Tr}\!\left(\e^{-\beta H_{\mathrm{SSH}}}O\right)
= \langle O\rangle.
\end{align}
Applying this to $T^{\mathrm{intra}}_0$ gives
\begin{align}
\langle \mathcal I T^{\mathrm{intra}}_0 \mathcal I^{-1} \rangle&= \mathrm{Tr}\!\left(\e^{-\beta H_{\mathrm{SSH}}}(T^{\mathrm{intra}}_0)^\dagger\right)\nonumber\\
&=\mathrm{Tr}\!\left(\e^{-\beta H_{\mathrm{SSH}}}T^{\mathrm{intra}}_0\right)^*\nonumber\\
&=\langle T^{\mathrm{intra}}_0\rangle^*.
\end{align}
Since inversion maps the operator onto itself spatially at the symmetric point, $\langle \mathcal I T^{\mathrm{intra}}_0 \mathcal I^{-1}\rangle=\langle T^{\mathrm{intra}}_0\rangle$, which implies 
\begin{align}
\langle T^{\mathrm{intra}}_0\rangle=\langle T^{\mathrm{intra}}_0\rangle^*.
\end{align}
So the expectation value of the intracell operator at the inversion center is real, if it is positive it has a phase which is zero, and if it is negative, the corresponding phase is equal to $\pi$.

In the strict topological limit ($t_1>0$, $t_2=0$), the closed expression in Eq.~\eqref{eq:Tintra-triv} gives 
\begin{align}
\langle T^{\mathrm{intra}}_0\rangle_{\rm{triv}}>0,
\end{align}
so the phase at the inversion-symmetric point is $\varphi^{\mathrm{intra}}=0$. 
The only way that the phase can change is if the expectation value becomes negative, by passing through zero. 
To verify that the expectation value never vanishes as $(t_1,t_2,\beta)$ vary (including at the transition $t_1=t_2$),the determinant representation
\begin{align}
\langle T^{\mathrm{intra}}_0\rangle
&=\det\!\big(1-f+f\,\e^{\i\delta k X^{\mathrm{intra}}_0}\big),\\
f&=(1+\e^{\beta H_{\mathrm{SSH}}})^{-1},
\end{align}
is expanded by treating the exponential perturbatively for small $\delta k=2\pi/N$.
This yields
\begin{align}
\langle T^{\mathrm{intra}}_0\rangle
&=\det\!\big[1+A+\mathcal O(\delta k^3)\big],
\end{align}
with
\begin{align}
A&:=f\!\left[\i\delta k X^{\mathrm{intra}}_0
-\tfrac{1}{2}(\delta k)^2 (X^{\mathrm{intra}}_0)^2\right].
\end{align}
Using the operator identity, $\ln\det(1+A)=\mathrm{Tr}\ln(1+A)$ and the series 
$\ln(1+A)=A-\tfrac{1}{2}A^2+\mathcal O(A^3)$ gives
\begin{align}
\label{eq:appendix_magnitudeTintra}
\ln\langle T^{\mathrm{intra}}_0\rangle
&=\i\delta k\mathrm{Tr}\!\big[f( X^{\mathrm{intra}}_0)\big]\nonumber\\
&-\frac{(\delta k)^2}{2}\,
\mathrm{Tr}\!\big[f(1-f)(X^{\mathrm{intra}}_0)^2\big]+\mathcal O((\delta k)^3).
\end{align}
The linear term vanishes by inversion symmetry at the inversion center:
\begin{align}
\Tr\!\big[f X^{\mathrm{intra}}_0\big]
&=\Tr\!\big[\mathcal I f X^{\mathrm{intra}}_0 \mathcal I^{-1}\big]\nonumber\\
&=-\,\Tr\!\big[f X^{\mathrm{intra}}_0\big].\nonumber
\end{align}
The quadratic term is real and nonnegative. 
To make this explicit, it is convenient to rewrite the trace in a manifestly positive form using the identity $\Tr[B^\dagger B]\ge0$ for any operator $B$. 
Define
\begin{align}
B=\sqrt{1-f}\,X^{\mathrm{intra}}_0\sqrt{f},
\end{align}
so that $\Tr[f(1-f)(X^{\mathrm{intra}}_0)^2]=\Tr[B^\dagger B]$. 
This construction requires that the operators $f$ and $(1-f)$ admit Hermitian square roots; $\sqrt{f}$ and $\sqrt{1-f}$ satisfying $(\sqrt{f})^2=f$ and $(\sqrt{1-f})^2=1-f$. 
Such roots exist because $f$ is positive semidefinite: its eigenvalues lie in the interval $[0,1]$, implying that both $f$ and $(1-f)$ are positive semidefinite operators. 
The operator $X^{\mathrm{intra}}_0$ is Hermitian, so $(X^{\mathrm{intra}}_0)^2$ is positive semidefinite, 
ensuring that $\Tr\,[f(1-f)(X^{\mathrm{intra}}_0)^2]$ is a nonnegative real number, vanishing only at zero temperature, where $f(1-f)=0$.

Exponentiating Eq.~\eqref{eq:appendix_magnitudeTintra} gives the magnitude 
\begin{align}
\Big|\langle T^{\mathrm{intra}}_0\rangle\Big|
&=\exp\!\left\lbrace-\tfrac{(\delta k)^2}{2}\,
\mathrm{Tr}\!\big[f(1-f)(X^{\mathrm{intra}}_0)^2\big]\right.\nonumber\\
&\left.\hspace{2cm}+\mathcal O((\delta k)^3)\right\rbrace.
\end{align}
The determinant is therefore strictly positive at all temperatures and values of hopping parameters.
Because $\langle T^{\mathrm{intra}}_0\rangle$ is real, its sign---fixed as positive in the strict trivial limit---cannot change.
The intracell operator at the inversion center thus has a positive expectation value and a vanishing phase, 
\begin{align}
\varphi^{\mathrm{intra}}=0.
\end{align}
The same argument holds for the value of the phase of the intercell operator at any inversion symmetric point.
The sign of the real expectation value is obtained from the analytic expression in the strict topological limit, which cannot change as the magnitude of the expectation value never becomes zero.
\bibliography{refs}
\end{document}